\documentclass[iop]{emulateapj}
\pdfoutput=1

\usepackage{rotating}
\usepackage{fontenc}

\slugcomment{To appear in APJ}

\shorttitle{Silicon carbide dust clusters}
\shortauthors{Gobrecht et al.}

\begin{document}


\title{Nucleation of small silicon carbide dust clusters in AGB stars}


\author{David Gobrecht\altaffilmark{1}, Sergio Cristallo \altaffilmark{1} and Luciano Piersanti \altaffilmark{1}}

\affil{Osservatorio Astronomico di Teramo, INAF, 64100 Teramo, Italy}
    
\and

\author{Stefan T. Bromley\altaffilmark{2,3}}
\affil{Departament de Ciència de Materials i Química Física and Institut de Química
Teòrica i Computacional (IQTCUB), Universitat de Barcelona, E-08028 Barcelona, Spain}
\affil{Institucio Catalana de Recerca i Estudis Avancats (ICREA), 08010 Barcelona, Spain}




\begin{abstract}
Silicon carbide (SiC) grains are a major dust component in carbon-rich AGB stars. The formation pathways of these 
grains are, however, not fully understood.\
We calculate ground states and energetically low-lying structures of (SiC)$_n$, $n=1,16$ clusters by means of simulated annealing (SA) and Monte 
Carlo simulations of seed structures and subsequent quantum-mechanical calculations on the density functional level of theory. We derive the 
infrared (IR) spectra of these clusters and compare the IR signatures to observational and laboratory data.\ 
According to energetic considerations, we evaluate the viability of SiC cluster growth at several densities and temperatures, characterising various 
locations and evolutionary states in circumstellar envelopes.\ 
We discover new, energetically low-lying structures for Si$_{4}$C$_{4}$, Si$_{5}$C$_{5}$, Si$_{15}$C$_{15}$ and Si$_{16}$C$_{16}$, and new ground states for Si$_{10}$C$_{10}$ and Si$_{15}$C$_{15}$.
The clusters with carbon-segregated substructures tend to be more stable by 4-9 eV than their bulk-like isomers with alternating Si-C bonds. 
However, we find ground states with cage (``bucky''-like) geometries for Si$_{12}$C$_{12}$ and Si$_{16}$C$_{16}$ and low-lying, stable cage structures for n $\ge$ 12. The latter findings indicate thus a regime of clusters sizes that differs from small clusters as well as from large-scale crystals. 
Thus, and owing to their stability and geometry, the latter clusters may mark a transition from a quantum-confined cluster regime to crystalline, solid bulk-material. 

The calculated vibrational IR spectra of the ground-state SiC clusters shows significant emission. 
They include the 10-13 $\mu$m wavelength range and the 11.3 $\mu$m feature inferred from laboratory measurements and observations, respectively, though the overall intensities are rather low.
\end{abstract}





\section{Introduction}

Dust is ubiquitous in the Universe and plays a crucial role in astrophysical environments. Dust impacts the synthesis of complex organic molecules 
in molecular clouds, the wind-driving of evolved stars and the formation of celestial bodies (e.g. asteroids, planets) in protoplanetary discs 
\citep{doi:10.1146/annurev.astro.38.1.427,1991A&A...248..105D,2014prpl.conf..339T}. 
Dust is thus essential for the chemical evolution of galaxies and its formation in late-type stars is the subject of this paper.
The dust formation from a gaseous medium requires several thermodynamic conditions: densities above a certain threshold to ensure sufficient 
collisions between the constituent particles, moderate temperatures below the stability threshold of the dust component, and sufficient time for the 
nucleation and growth of molecular clusters into larger grains. Such conditions are found in the warm and dense molecular layers in circumstellar 
environments of Asymptotic Giant 
Branch (AGB) stars \citep{1999A&A...348L..17W}. It is thus not surprising, that, among the stellar sources of dust, AGB stars are a significant 
contributor. We note, however, that, arguably, the bulk of dust present in the local Universe could be the result of grain growth and reprocessing in the interstellar medium  \citep{2009ASPC..414..453D}.

The amount and nature of the dust depends on stellar mass, metallicity, and not least on the photospheric C/O ratio.
For C/O \textless 1 (M-type AGB stars) the circumstellar chemistry is oxygen-dominated and the type of dust that is forming is made of silicates, 
alumina and other metal oxides \citep{2016A&A...585A...6G}. In carbon-rich stars with C/O \textgreater 1, carbonaceous molecules prevail and condensates such as 
amorphous carbon and silicon carbide constitute the dust grains \citep{2010LNP...815.....H}. 
About 90\% of SiC grains are thought to come from low-mass AGB stars of approximately solar metallicity \citep{Davis29112011} and 
SiC accounts for about $\sim$ 10\% of carbonaceous dust of solar and moderately subsolar metallicity \citep{2013A&A...555A..99Z}.
 
S-type AGB stars are reckoned as transitional objects between M-type and C-type stars, respectively, and have little 
excess of either carbon or oxygen. However, these stars may produce dust in the form of pure metals \citep{2006A&A...447..553F}.

Often, it is argued that dust formation and the related mass loss phenomena is less understood in M-type stars, owing to the low opacity of oxygen-
rich condensates in the near infrared range \citep{2006A&A...460L...9W,2008A&A...491L...1H}. The wind-driving in carbon-rich AGB atmospheres is 
better understood. However, the synthesis of carbonaceous dust clusters and the formation routes towards (silicon)-carbon grains is not yet fully understood.
 
One of the major dust components is silicon carbide (SiC) showing a spectral emission/absorption in the 10-13 $\mu$m range, in particular a strong 
and characteristic feature around $\sim$ 11.3 $\mu$m \citep{1996NASCP3343...61S,2006ApJ...650..892S}.
Laboratory studies have shown that the spectral band profile depends on size, shape and purity, respectively, of the SiC dust grains, but is less 
affected by its crystal type \citep{1999A&A...345..187M}. Fundamental lattice vibrations (i.e. phonons) dominate the interaction of infrared 
radiation with crystalline SiC. For small dust clusters, however, the situation is different. Owing to the lack of periodicity and lattices it is 
impossible to excite collective lattice vibrations such as phonons in small clusters. However, clusters possess distinct and non-bulk-like 
vibrational and rotational modes arising due to 
bending and stretching of internal bonds.
   
Moreover, SiC dust grains have been found in primitive meteorites \citep{1987Natur.330..728B} and have typical sizes of 0.3-3 
$\mu$m \citep{1994GeCoA..58..459A}. 
The analysis of the SiC grain isotope composition, in particular the excess of $^{13}$C and $^{15}$N compared with scaled-solar abundances, 
revealed unambiguously that SiC grains originate from extended atmospheres of AGB stars \citep{HOPPE1996883}. In addition, the 
majority of the SiC grains exhibit s-process isotopic signatures arising in the atmospheres of Carbon-rich AGB stars 
\citep{1994ApJ...430..858G,0004-637X-786-1-66,0004-637X-803-1-12}.
More recently, new instruments like the NanoSIMS \citep{2007GeCoA..71.4786Z} became available and allowed the analysis of SiC grains with sizes as 
small as a few nm \citep{2009LPI....40.1198H}.
The investigation and analysis of rather small SiC grains (0.2$\mu$m - 0.5$\mu$m) have revealed that submicrometer-sized grains originating from 
AGB stars are much more abundant than their larger, micron-sized counterparts \citep{2010ApJ...719.1370H,2014AIPC.1594..307A}.

The classification of individual SiC grains into different groups named ``Mainstream'', AB and X grains, respectively, is based on the isotopic 
excess. ``Mainstream'' grains are associated with carbon-rich AGB stars. About 90\% of the presolar SiC grains are thus thought to come from low-
mass AGB stars of approximately solar metallicity \citep{Davis29112011}. 

Owing to the interaction with the stellar radiation field, SiC grains are promising candidates to trigger the mass-loss in carbon-rich AGB stars.
However, the formation of the SiC grains in stellar winds remains poorly understood.    

In the bulk phase, SiC exists in about 250 crystalline forms, so called polytypes. The most commonly encountered polytypes are $\alpha$-SiC and 
$\beta$-SiC with tetrahedrally coordinated Si atoms. All SiC grains extracted from meteorites have proven to be either cubic $\beta$(3C)-SiC 
($\sim$ 80 \%) or hexagonal $\alpha$(2H)-SiC \citep{2003GeCoA..67.4743D,2005ApJ...631..988B}. These two polytypes do not 
differ systematically in their spectral signatures \citep{1999A&A...345..187M}. The band profile is rather affected by grain size, shape, and 
impurities, respectively.
Moreover, the analysed SiC grains do not contain any seed nuclei of different chemical type in 
their centres \citep{2005LPI....36.2010S}, thus indicating a homogeneous (homo-molecular) grain formation.
The properties of nanoparticles with sizes below 50 nm, however, differ significantly from bulk properties. Quantum and surface effects of these 
small particles lead to non-crystalline structures, whose characteristics (geometry, coordination, density, binding energy) may differ by orders 
of magnitude, as compared to the bulk material. 
In the smallest clusters, namely dimers and polymers of a dust species, the inter-atomic bonds are often unsaturated (in terms of atomic 
coordination), owing to the high surface-to-volume ratio. A top-down approach, i.e. deducing cluster characteristics from bulk material properties, 
is thus inappropriate.  
Contrary, a bottom-up approach, starting with prevalent molecules in the gas phase (e.g. SiC, SiO) and successive growth to clusters by molecular 
(addition) reactions, seems to be suitable. 
Such a method has been applied for clusters of magnesium oxide \citep{1997A&A...320..553K}, titanium dioxide \citep{2015A&A...575A..11L}, silicates of enstatite and forsterite stoichometry \citep{2012MNRAS.420.3344G}, and silicon oxide \citep{C6CP03629E}, respectively.   

In Section II we describe the computational methods used to characterise the SiC cluster structures and energetics. Section III 
gathers and summarises the results for the most stable clusters. Finally, the results are discussed with particular attention on implications for 
circumstellar dust formation and spectroscopic signatures.   

\section{Methods}


In this study, global optimisation techniques and Molecular Dynamic (MD) simulations are used to determine the energetically most stable cluster structures.
The more atoms a cluster contains, the larger its size is and the number of possible structural isomers increases drastically. The investigation of 
large clusters is therefore computationally demanding.   
In order to reduce the computational effort, we apply several pre-selection methods to find potential minimum energy SiC cluster structures.
Seed cluster structures are constructed by hand according to their geometries reported in the literature.

\subsection{Monte-Carlo Basin-Hopping search on the Buckingham potential energy landscape} 
Some of the candidate structures are found with the Monte Carlo - Basin Hopping (MC-BH) 
global optimization technique \citep{doi:10.1021/jp970984n} with inter-atomic Buckingham pair potentials. The general form of the inter-atomic Buckingham pair potential reads:

\begin{equation}
U(r_{ij}) =  \frac{q_{i}q_{j}}{r} + A\exp(-\frac{r_{ij}}{B}) - \frac{C}{r_{ij}^6} 
\label{buck}
\end{equation}

\noindent where r$_{ij}$ is the relative distance of two atoms, q$_i$ and q$_j$ the charges of atom i and j, respectively and A, B and C the 
Buckingham pair parameters.
The first term represents the Coulomb law, the second term the short-range, steric repulsion term accounting for the Pauli principle, and the last 
term describes the van-der-Waals interaction. The steric repulsion term is motivated by the fact that atoms are not dot-like but occupy a certain 
volume in space.

In the case of silicon carbide, parameter sets for the Si-C system are lacking in the literature for several reasons.
As an integral part the electrostatic Coulomb potential appears in Equation \ref{buck}. It describes the repulsion and attraction of charged 
particles, in this case of the silicon and carbon ions within a SiC cluster. As lightest Group IV elements in the periodic table, Si and C form 
strong covalent bonds. The electronegativity (EN) of carbon (EN(C) = 2.55) is too small to allow carbon to form C$^{4-}$ or C$^{4+}$ ions.
The Buckingham potential is thus mainly used for materials with an ionic character and as for 
example metal oxides.
Nevertheless, there is a significant amount of charge transfer of 2.5 electrons between Si and C atoms \citep{B902603G}.  
Nonetheless, \cite{B902603G} have shown the similarity of zincblende ZnO (a cubic crystal type with face-centred lattice points), and $\beta$ SiC, 
despite the first is generally regarded as ionic II-VI system and the latter as covalent IV-IV system. 
Moreover, they found that the Buckingham parameters for ZnO also describe SiC clusters fairly well. We therefore performed 
MC-BH with a simplified version of the parameter set for ZnO given by \cite{whitmore2002surface}.

The ZnO forcefield we employ has been shown to be able to stabilize a wide range of different cluster isomers 
\citep{doi:10.1021/jp805983g} and bulk polymorphs 
\citep{C3NR04028C} which exhibit alternating cation-anion ordering..
However, to reduce the probability to miss stable cluster isomers in our searches, we also ran some test calculations for several sizes 
with a forcefield parameterized for ZnS \citep{JM9950502037} which potentially provides an additional source of cluster isomers not easily 
found with the ZnO forcefield. However, the few structures that we found exclusively with the ZnS parameters had high energies 
(when converted to SiC clusters) and did not compete with the ZnO cluster analogues. Although the use of forcefields is an approximation, 
their use enables us to perform tractable thorough searches. With our mixed-forcefield approach (see also section 2.2) 
we hope to have minimized the probability to miss a stable SiC isomer.


\subsection{Tersoff potential simulated annealing}
As already explained, the Buckingham pair potential may fail to describe stable cluster configurations, which show segregation of the Si and C atoms.
In this case, the stable clusters are characterised by rather covalent than ionic bonds. 
A simple two-body interaction is thus not sufficient to properly describe the Si-C system.
In addition, a three-body potential is needed to describe the covalent character of bond bending and stretching
\citep{PhysRevB.31.5262,:/content/aip/journal/jap/101/10/10.1063/1.2724570}.
In order to properly describe internal interactions of the most stable SiC clusters, empirical bond-order 
potentials are favourable, in particular for small clusters \citep{PhysRevB.71.035211}.  This class of interatomic potentials include the Tersoff-
type \citep{PhysRevB.39.5566}, the Brenner \citep{1990PhRvB..42.9458B}, or, ReaxFF 
\citep{doi:10.1021/jp004368u}, which take into account the bonding environment, namely the bond length, the angle and the number of bonds. As a 
consequence of 
geometry,  the bonding angle in a tetrahedrally coordinated system like SiC is $\Theta$ = arccos(-1/3) = 109.47$^{\circ}$. The general form of a 
bond-order potential reads:

\begin{equation}
V(r_{ij}) = f_{c}(r_{ij}) \left[V_{rep}(r_{ij}) + b_{ij}V_{att}(r_{ij})\right]
\end{equation}

\noindent where  V$_{rep}(r_{ij})$ = A$_{ij}\exp{-\lambda_{ij} r_{ij}}$ is the repulsive part of the potential and V$_{att}(r_{ij})$=B$_{ij}\exp{-\mu_{ij} r_{ij}}$ the attractive effective potential.
b$_{ij}$ modifies the strength of the bond, depending on the environmental parameter like the bonding angle $\Theta$ as reported in \cite{PhysRevB.39.5566}.
In the Tersoff parametrisation of inter-atomic Si-C molecular system, which is chosen in our approach, the potential is modified by a taper function f$_c$.
f$_c$ is 1 for inter-atomic distances r$_{ij}$ smaller or equal of typical bonding distances and falls quickly to 0 for distances larger than $S$ and thus restricts the interaction to the first neighbouring atoms within a distance S.

\begin{equation}
f_c (r_{ij})= \left\{\begin{array}{ll} 1, & r_{ij} < R \\
         0.5 + 0.5 \cos(\frac{\pi(r_{ij}-R)}{S-R}), & R < r_{ij} < S \\
         0, & r_{ij} > S \end{array}\right. .
\end{equation}

The parameter set given by \cite{PhysRevB.39.5566} suffers from an underestimation of the dimer binding energy and may not be satisfactory for the 
description of small gas-phase molecules and clusters. A revised set of parameters is available \citep{PhysRevB.71.035211}.
In the updated parameter set, the bond-order term b$_{ij}$ is formulated differently from the original description.
Unfortunately, the new formulation is not compatible for calculations in most molecular dynamics programs.
However, the classic Tersoff parametrisation is sufficient for our purposes, as the results are subsequently refined using a quantum mechanical level of theory.
We use the programme GULP (General Utility Lattice Programme) \citep{A606455H} which is taylored for the classic parametrisation by \cite{PhysRevB.39.5566}.\\
Some SiC cluster structures have been reported in the literature 
\citep{2004physics...8016P,:/content/aip/journal/jcp/128/15/10.1063/1.2895051,molecules18078591}. We 
tested their stability against (small) distortions in molecular dynamics runs with GULP. Furthermore, 
we applied the classic Tersoff potential to these structures. In the majority of the cases, this potential suffices to stabilise the structures. In 
some cases, however,
the Tersoff potential fails to stabilize the clusters, and hand-constructed structures were taken instead for the subsequent computation. In some of these failure cases new, unreported clusters appeared.\

We also perform simulated annealing runs using the Tersoff-optimised structures - an imitation of a cluster cooling process.
The melting point, where crystalline SiC decomposes, is around  3000 K, which is chosen to be the maximal temperature in the annealing routine.   
By varying the starting temperature T$_{max}$ and the cooling timescales, we performed several hundred simulated annealing runs for the previously defined 
seed cluster structures. All the structures were cooled to a final temperature of 200 K. We distinguished between four regimes:
\begin{itemize}

\item High temperature annealing with T$_{max}$=3000 K
\item Moderate temperature annealing with T$_{max}$=1800 K
\item Low temperature annealing with T$_{max}$=1000 K
\item Molecular dynamics at a constant temperature of 300 K
\end{itemize}

The majority of the investigated clusters already stabilize around 600-800 K.  
In order to reinforce the convergence of the MD runs, the structures were optimised to the Tersoff potential at every step where the temperature is 
decreased. In the MD runs at constant temperature, snapshots of the lowest potential energy configurations were selected and further inspected.  

\subsection{Quantum-mechanical refinement}
Once pre-optimised, the clusters are refined using quantum-mechanical DFT (Density Functional Theory) calculations to obtain structure-specific infrared spectra (i.e. vibrational frequencies) rotational constants, and zero-point-energies.\  
By comparing the obtained infrared spectra with observational data, the specific isomers present in circumstellar envelopes can thus be identified.
The (SiC)$_n$ cluster structures, so far reported in the literature, rely on various theoretical quantum chemistry methods.
They include DFT methods using generalized gradient approximation (GGA,PBE), local density approximation (LDA), B3LYP and M11 functionals, 
respectively, and post-Hartree Fock methods using M\o ller-Plesset (MP2, MBPT) and coupled-cluster (LCCD, CCSD) techniques.
For DFT methods the computational cost scales with the system size as between the order $\mathcal{O}(N^3)$ and $\mathcal{O}(N^4)$, where $N$
is the number of electrons in the cluster. This means that they can be readily applied to systems containing 10s of atoms. However, many DFT methods can suffer from artificial electron self-interaction that results in overly strong electron delocalisation and too low potential energies
. In contrast, Post-Hartree–Fock methods do not suffer from these effects. However, the computational cost of these latter methods is very high and scales with the system size as $\mathcal{O}(N^5)$-$\mathcal{O}(N^7)$. 
They are thus prohibitive for systems of more than approximately 10 atoms. Functionals such as B3LYP and M11 attempt to compensate for the above mentioned shortcomings of typical GGA/LDA functionals. The recent extensive benchmark study by 
\cite{:/content/aip/journal/jcp/145/2/10.1063/1.4955196} confirms that the M11 functional is able to correctly identify all investigated (SiC)$_n$ ground states. Although B3LYP was found to be less accurate than M11 for SiC clusters, we also include data calculated with this widely used functional for comparison. We conclude that, for our purposes, the M11 functional method is the best compromise between a reasonable computational cost and the required accuracy.

Owing to its high computational costs, DFT calculations are performed on supercomputers using the well-
parallelised code \textit{Gaussian 09}. 
These calculations approximate the wave functions and the energy of a 
quantum many-body system in a stationary 
state. In the case of SiC clusters, the hybrid B3LYP functional with a cc-pVTZ (correlation-
consistent polarized Valence Triple Zeta) basis set is used \citep{1993JChPh..98.1372B}. Recent investigations, 
however, revealed that the B3LYP functional may fail to predict the correct ground states and spacings in 
relative energies for SiC clusters \citep{:/content/aip/journal/jcp/145/2/10.1063/1.4955196}. Moreover, the authors indicated in 
their benchmark study that the Minnesota functionals (e.g. M11, \cite{doi:10.1021/jz201525m}) have a more adequate accuracy, compared with B3LYP. 
Owing to this reason we additionally performed DFT calculations using the M11 functional for the majority of the investigated clusters.

\textit{Gaussian 09} optimises cluster structures at standard conditions (i.e pressure of 1 atm and temperature of 298 K). 
In circumstellar envelopes, however, very different conditions prevail: pressures are 4-9 orders of magnitude lower and temperatures factors of $\sim$ 5-200 higher.
In order to account for the above mentioned temperatures and pressures, the thermodynamic potential functions (enthalpy, entropy, Gibbs energy) are 
evaluated with the help of partition functions. These functions and their derivatives are calculated from the electronic energies, moments of 
inertia and vibrational frequencies within the rigid-rotor harmonic oscillator approximation
\citep{mcquarrie1999molecular,2012MNRAS.420.3344G}.  

As a consequence, the relative energy spacings of the individual clusters shift and may cross. This implies that the initial     
lowest energy isomer may not be the most favourable structure in circumstellar conditions and a different cluster structure is preferred. It is 
thus necessary to study a range of the energetically lowest-lying structures for each cluster size.  
The use of partition functions relies on the validity of thermodynamic equilibrium. We note, however, that AGB atmospheres may depart from equilibrium as they are periodically crossed by pulsational shock waves. The resulting Gibbs Free energies thus have limited validity.
Nonetheless, they provide a good approximation for the individual cluster stability in circumstellar conditions.




\section{Results}
In this Section we describe our results on the (SiC)$_n$,n=1-16, clusters. 
We constrain our calculations to a maximum size of n $\le$ 16. On one hand the DFT calculations rapidly become increasingly costly 
with increasing size. On the other hand, we follow a bottom-up approach thus focusing on the initial steps of SiC dust nucleation. These steps often represent the bottleneck of cluster nucleation processes.
The displayed numbers correspond to 
values obtained with the M11 functional, whereas the values in parenthesis correspond to B3LYP results.\\
\textbf{SiC}: As a diatomic molecule the SiC monomer is a linear structure. The SiC triplet represent 
the ground state of this molecule and is 1.34 (0.96) eV lower in energy than the corresponding singlet state.  
We find an average bond length of 1.707 (1.813) \AA , a rotational constant of 20.6 (18.3) GHz and a vibrational 
frequency of 1008.7 (862.2) cm$^{-1}$, which corresponds to a wavelength of 9.9 (11.6) $\mu$m.\

\begin{figure}[h!]
\plottwo{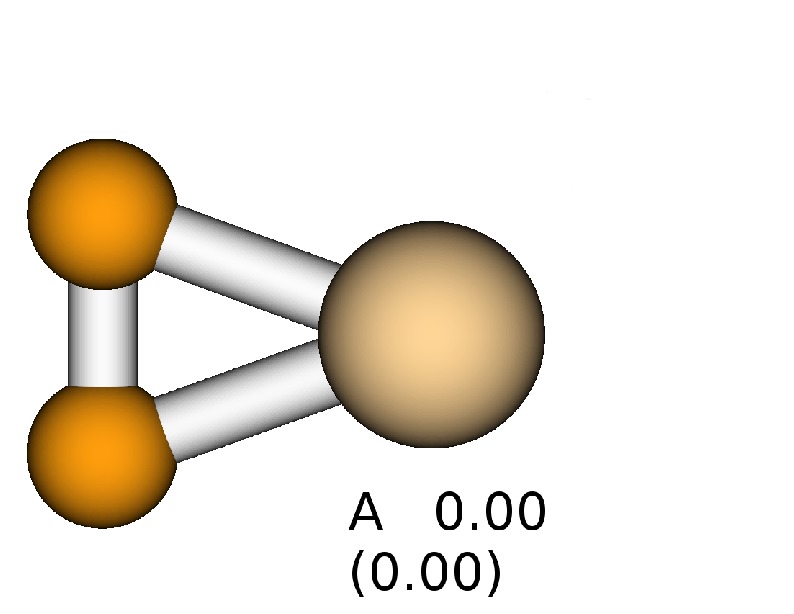}{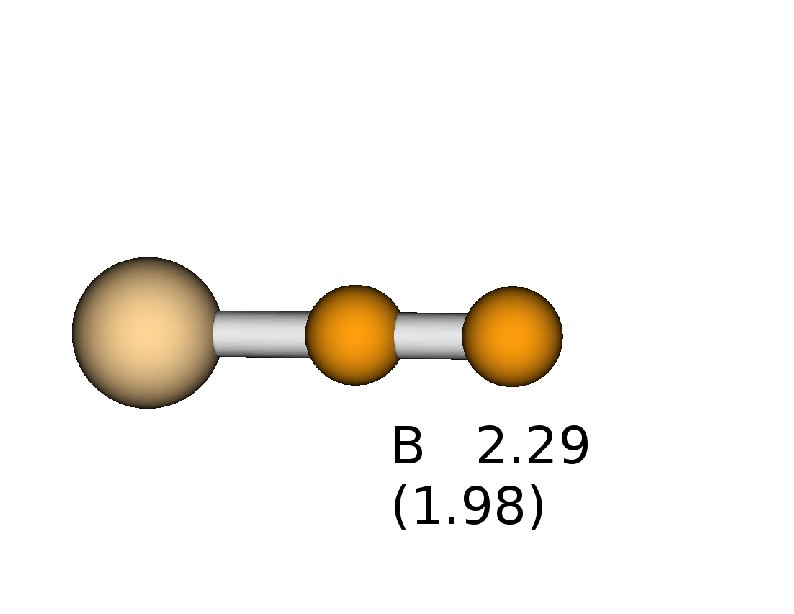}
\epsscale{0.5}

\plotone{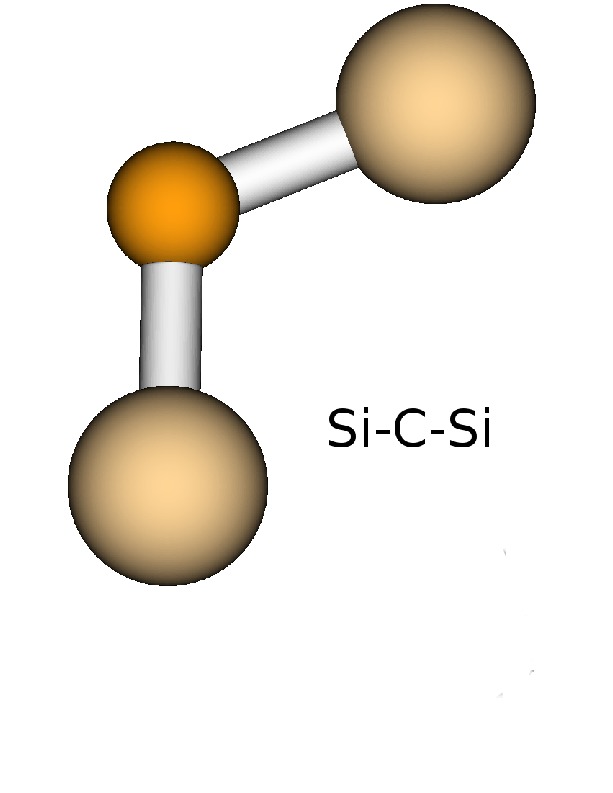}
\caption{The three stable SiC$_2$ clusters and the Si$_2$C ground state with relative energies (in eV)\label{fig1}.}
\end{figure}

\textbf{Si$_2$C and SiC$_2$}:\
The SiC$_2$ ground state is a triangle and lower by 2.29 (1.98) eV than the linear triplet isomers B in 
Figure \ref{fig1}. The isomer with a linear C-Si-C chain is unstable (6.82 eV above the ground state). 
Regarding the large differences in energy, we assume that isomer A is the dominant state of SiC$_2$ and the 
geometry of B is negligible for all temperatures and pressures. 

\cite{:/content/aip/journal/jcp/142/23/10.1063/1.4922651} have characterized the ground electronic state of 
Si$_{2}$C.
The singlet isomer with two off-axis Si atoms bended by an angle of 114.87 deg. and a C$_{2v}$ symmetry, 
reported in \cite{doi:10.1021/acs.jpclett.5b00770,2015ApJ...806L...3C}, is 
found by our M11 calculations, but not with the B3LYP functional. In the latter case, the molecule relaxes into 
linear C-Si-C or fails to converge. This result demonstrates the advantage of the M11 functional, compared with 
B3LYP.    
The linear Si-Si-C isomer exhibits imaginary frequencies in their IR spectra. Structures showing imaginary frequencies (vibration modes) represent 
a saddle point (and not a minimum) in the complex potential energy landscape. These saddle points have, as the real minima, a zero gradient and are 
interpreted as transition states. 

Thus, the bended C-Si-C structure is the only stable cluster we report for Si$_{2}$C.\

\begin{figure}[h!]
\plottwo{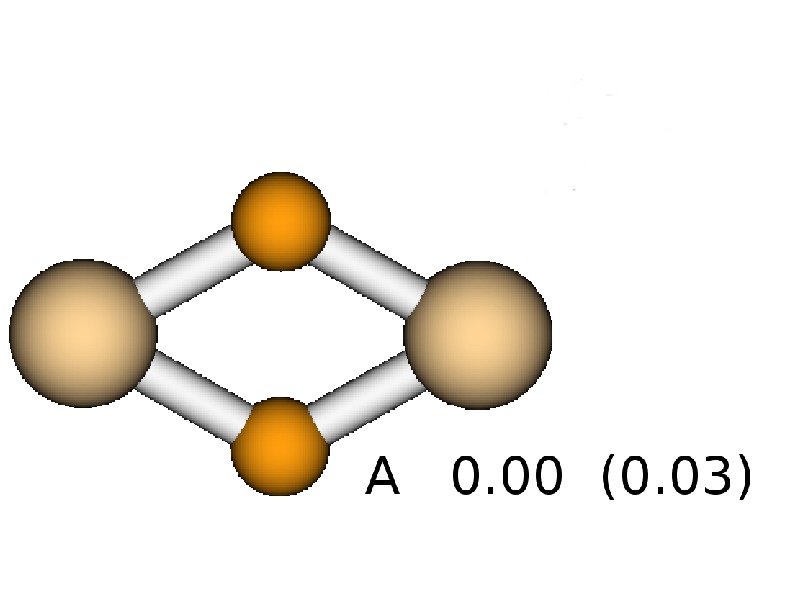}{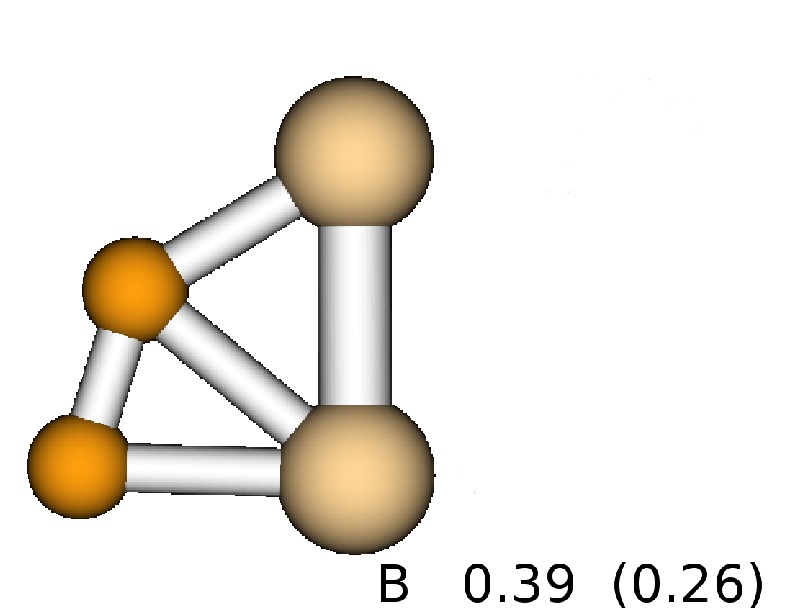}
\plottwo{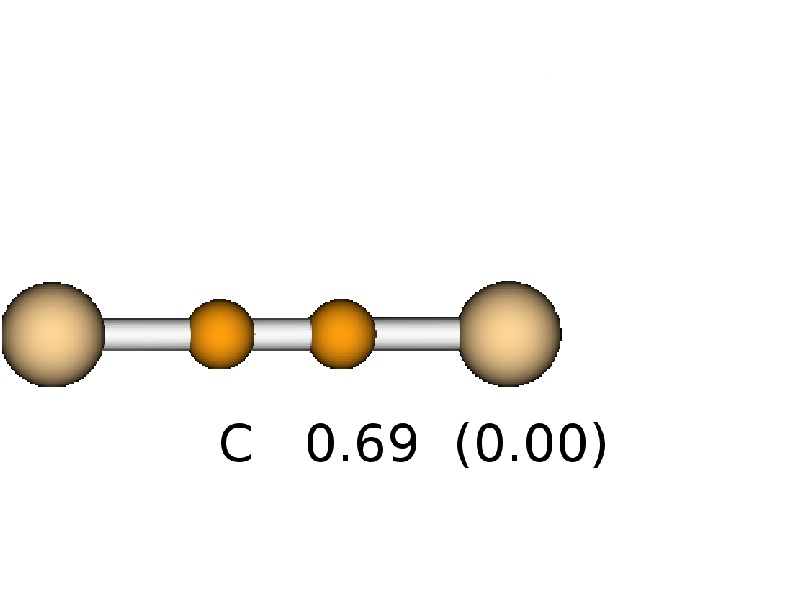}{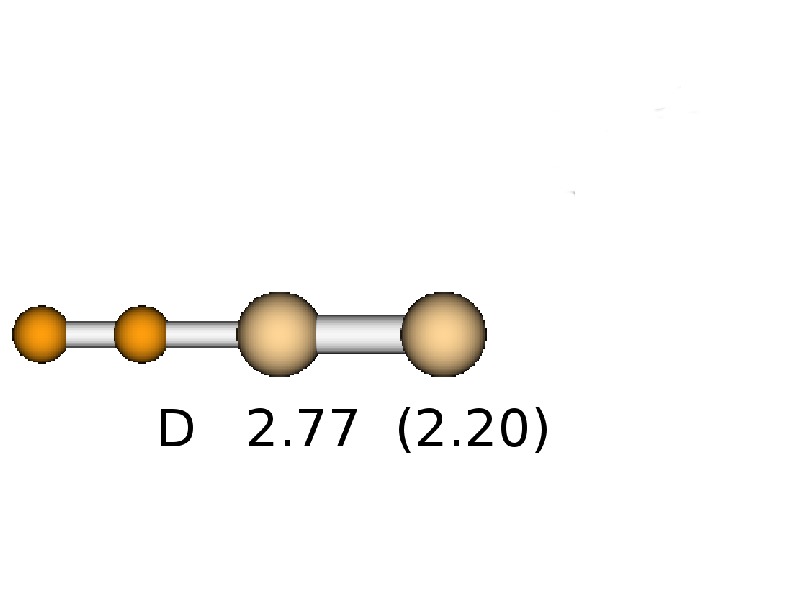}
\caption{The most stable Si$_2$C$_2$ clusters and relative energies (in eV).\label{fig2}}
\end{figure}

\begin{figure}[h!]
\plottwo{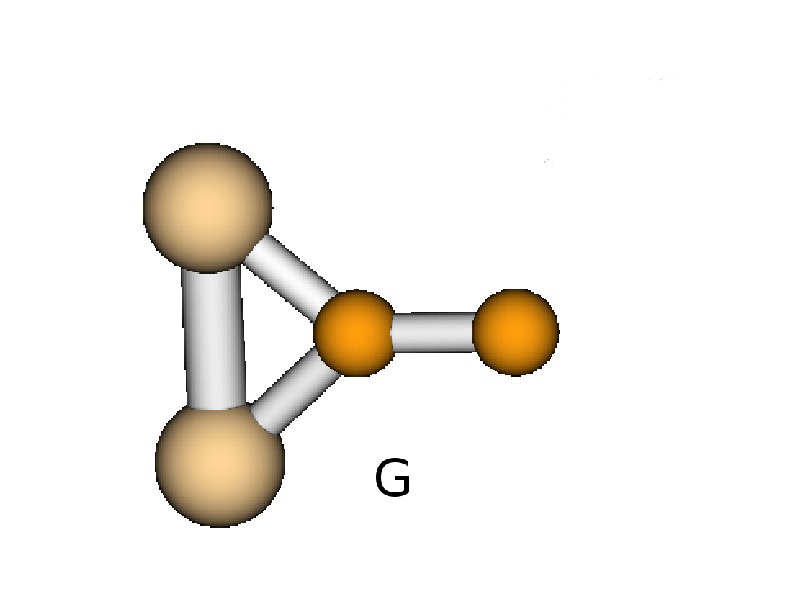}{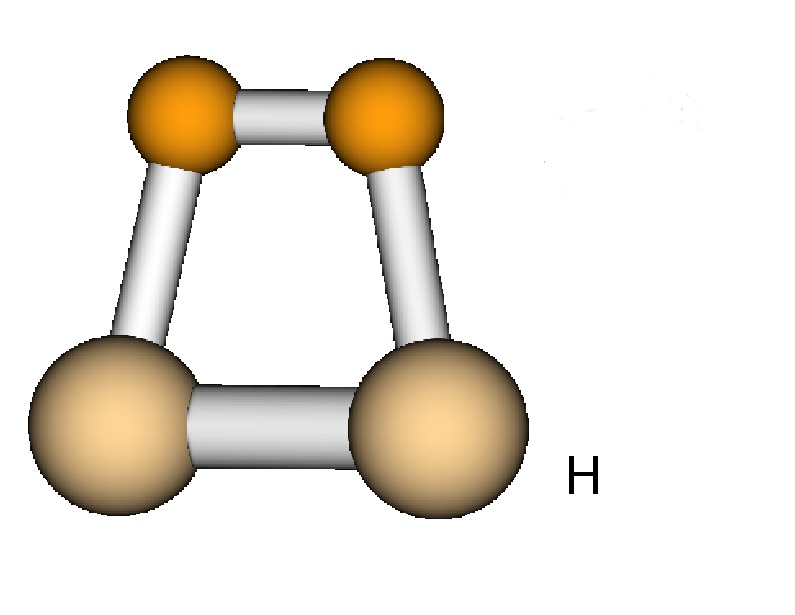}
\epsscale{0.5}
\plotone{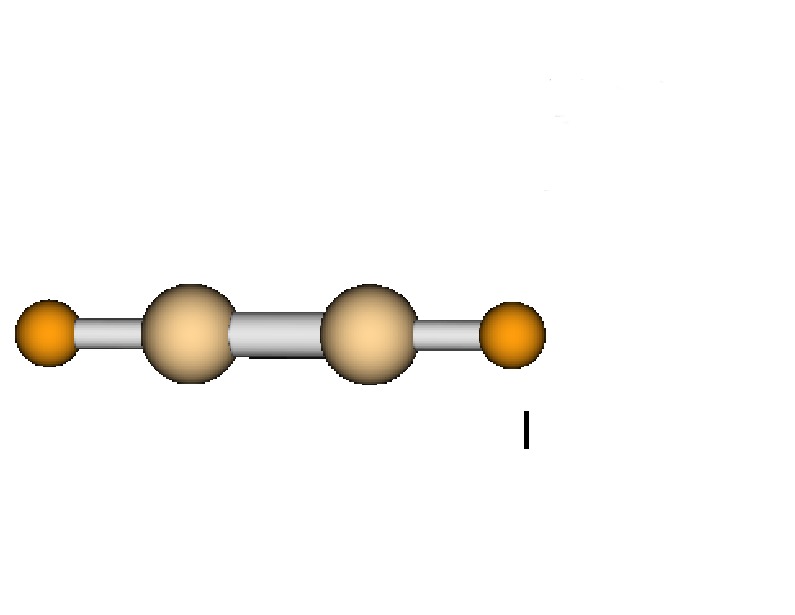}
\caption{Transition states of Si$_2$C$_2$ clusters.\label{fig2b}}
\end{figure}

\textbf{Si$_{2}$C$_{2}$}:\ 
Two structures of Si$_2$C$_2$ are commonly proposed as ground states: the linear triplet structure and the closed 
rhomb. They usually show a tiny difference in binding energies and are thus considered as degenerate isomers. The 
exact energy separation depends on the used functional and basis set. This is consistent with our B3LYP findings, 
where these two structures are separated by only 0.03 eV, as can be seen in Figure \ref{fig2}. 
Contrary, we show that the M11 functional predicts the rhombic structure (A) to be more stable than 
the linear chain (C) by 0.69 eV at standard conditions. 
Isomer B is characterized by threefold-coordinated (Si and C) atoms and is 0.39 (0.26) ev above the lowest-lying 
state.
The linear triplet structure D has potential energy 2.77 (2.20) eV higher with respect 
to the ground state. The structures shown here have been previously found by \cite{2004physics...8016P}, 
\cite{:/content/aip/journal/jcp/128/15/10.1063/1.2895051}, and \cite{Duan2010630}, respectively. Several further 
isomers have been investigated by 
\cite{2004physics...8016P}. Our calculations show, however, that the structures G, H and I in Figure \ref{fig2b} 
are transition states and the force constants indicate a relaxation into isomer B of Figure \ref{fig2}.\

\begin{figure}[h!]
\plottwo{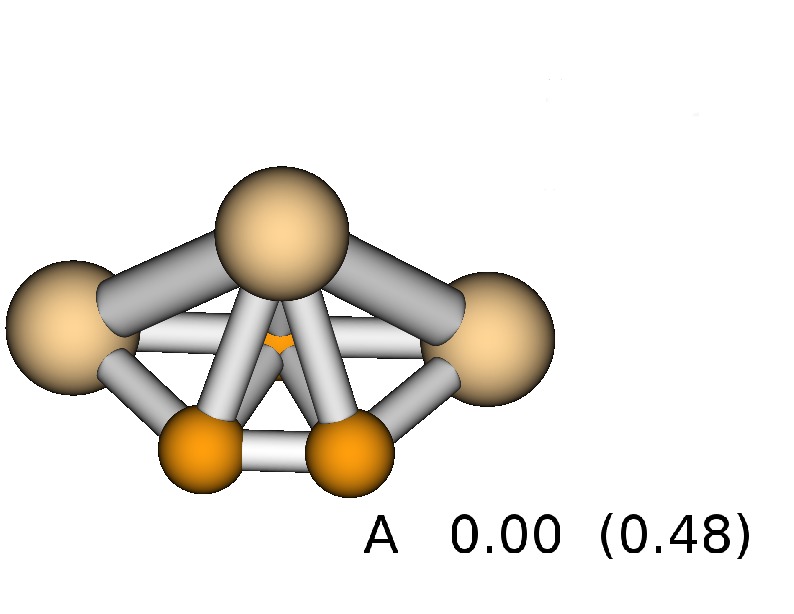}{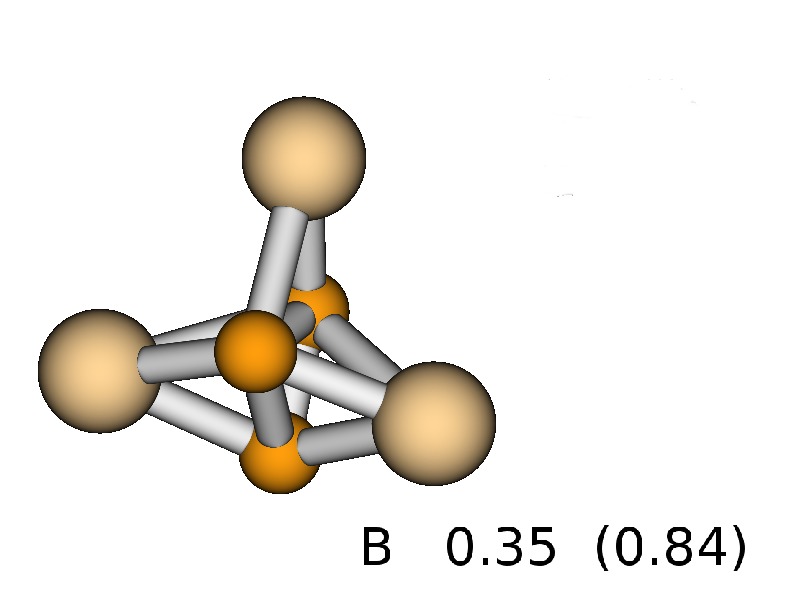} 
\plottwo{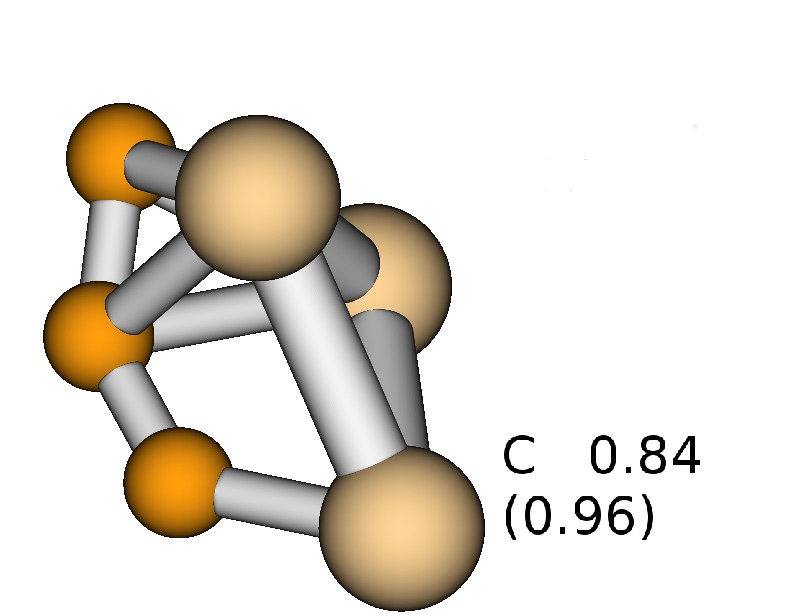}{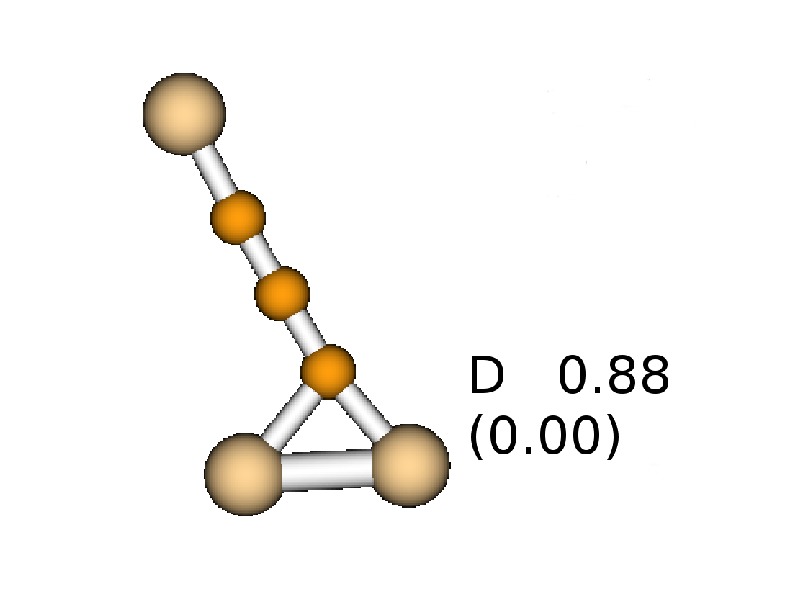}
\caption{The most stable Si$_3$C$_3$ clusters and relative energies (in eV).\label{fig3}}
\end{figure}

\textbf{Si$_{3}$C$_3$}:\
For Si$_3$C$_3$ clusters an extensive and comprehensible study has been carried out by \cite{Muhlhauser1994} 
who examined 17 structure in total. Further studies by 
\cite{:/content/aip/journal/jcp/128/15/10.1063/1.2895051,2004physics...8016P,Duan2010630} have revealed that some of these structures are particularly stable. In Figure \ref{fig3}, we summarise our findings.\

\begin{figure}[h!]
\plottwo{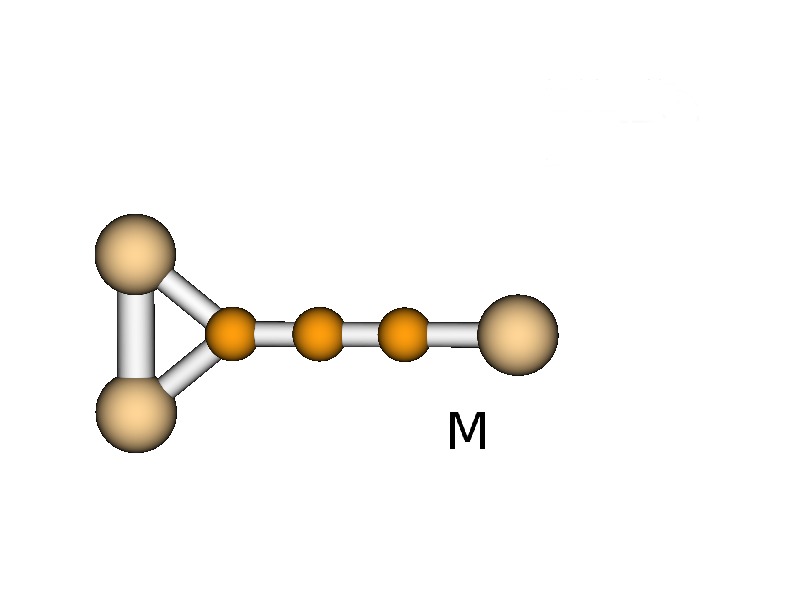}{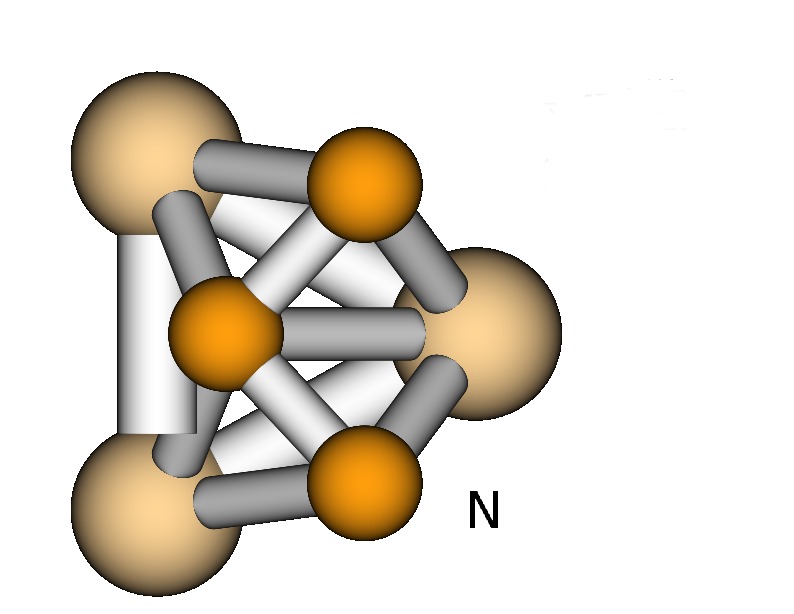}
\epsscale{0.5}
\plotone{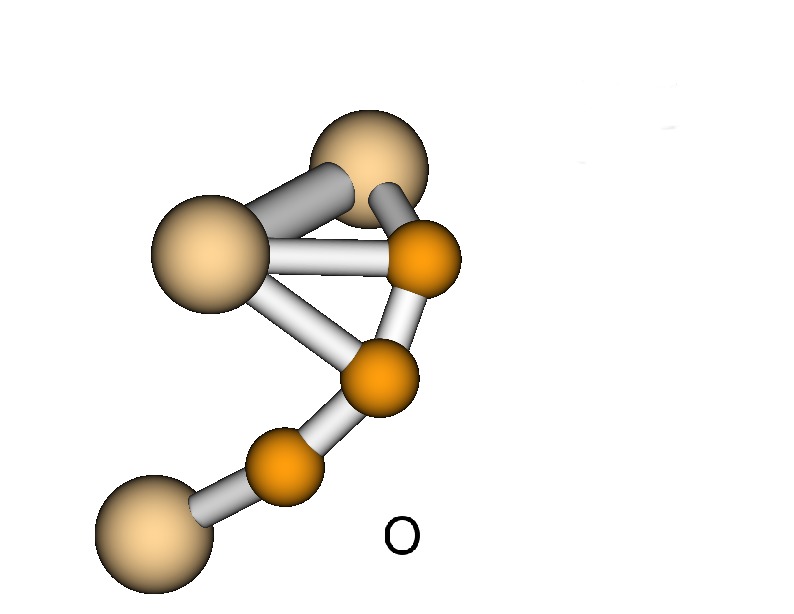}
\caption{Transition states of Si$_3$C$_3$ clusters.\label{fig3b}}
\end{figure}

Apparently, the isomers in Figure \ref{fig3} are characterised by planar structures as well as three-dimensional 
forms with triangular faces. 
All clusters contain three adjacent C atoms. 
The majority of the found Si$_3$C$_3$ clusters show a carbon chain (like C and D)
but also triangular C arrangement is observed (A and B). 
Ground state A and next higher-lying isomer B are non-planar and 
have triangular faces. Structure D is the lowest-lying isomer within the B3LYP level of theory. In the M11 
functional frame, however, D is 0.88 eV above the minimum structure and the C$_3$ chain is slightly bent.
Some structures that have been reported in the literature, however, exhibit imaginary infrared frequencies indicating a transition state rather than a 
minimum structure. By identifying the 
bond causing the imaginary vibration and re-optimising a slightly distorted structure, we found that structures 
M, N, and O in Figure \ref{fig3b} 
relax into 
other low-lying structures. All linear structures are triplet states and are energetically unfavourable  or exhibit 
imaginary frequencies and can thus not be considered as minimum structures.

\begin{figure}[h!]
\plottwo{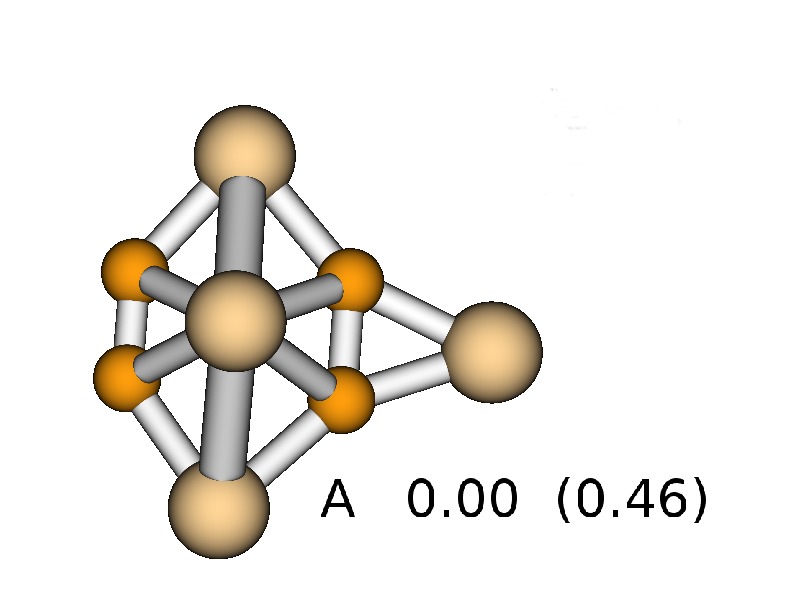}{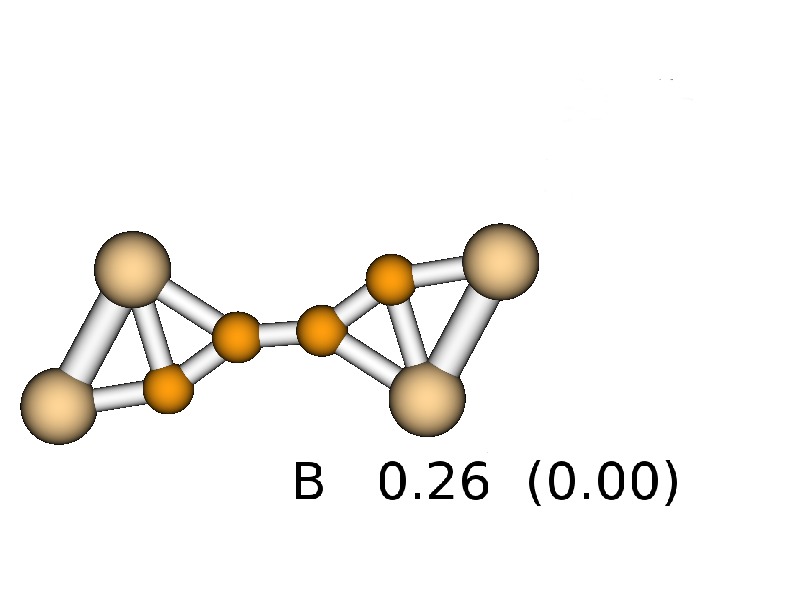}
\plottwo{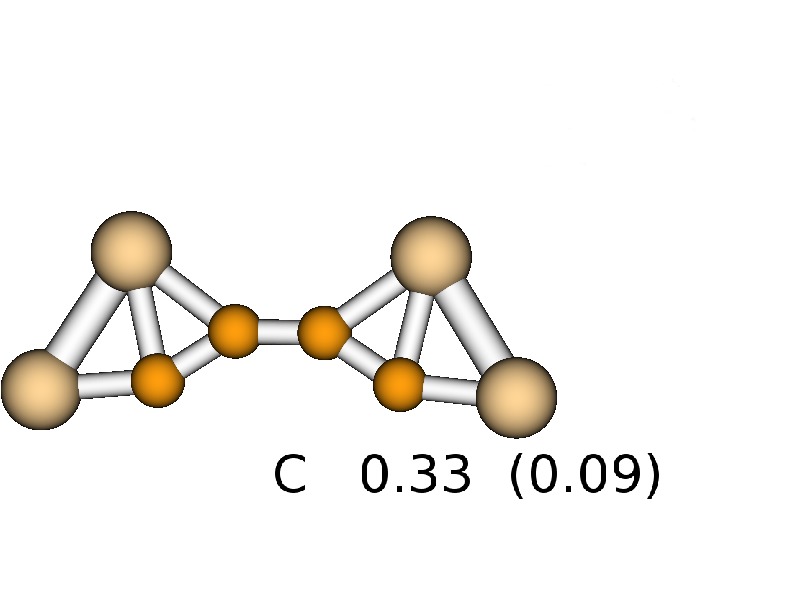}{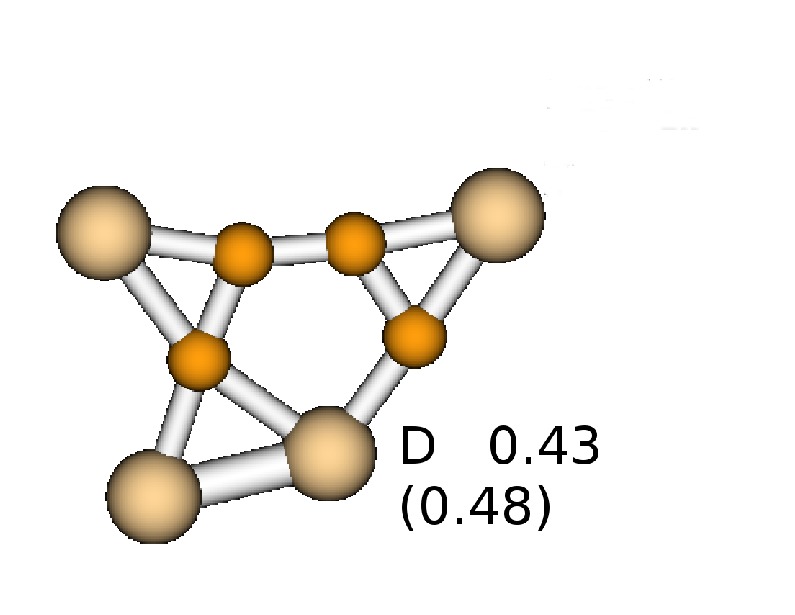}
\caption{The most stable Si$_4$C$_4$ clusters and relative energies (in eV).\label{fig4}}
\end{figure}

\textbf{Si$_{4}$C$_4$}:\ 
The ground state (A) of Si$_{4}$C$_{4}$ displayed in Figure \ref{fig4} is a non-planar structure having one Si 
atom out of the plane. The second lowest energy structure (B) is a planar structure with a C$_{2h}$ symmetry. It is composed 
of a 4-member-trans-carbon chain and can be viewed as two connected Si$_2$C$_2$ clusters (isomer B in Figure 
\ref{fig2}), bridged by C-C bonding. The 
corresponding cis-isomer (structure C) has an energy 0.33 eV above the ground state and 0.07 (0.09) above the trans-isomer. Trans- 
and cis-isomers differ only by a rotation of 180$^{\circ}$ along the C-C double bond axis. 
The structures A-D have been reported in \cite{molecules18078591} and references therein. 

\begin{figure}[h!]
\plottwo{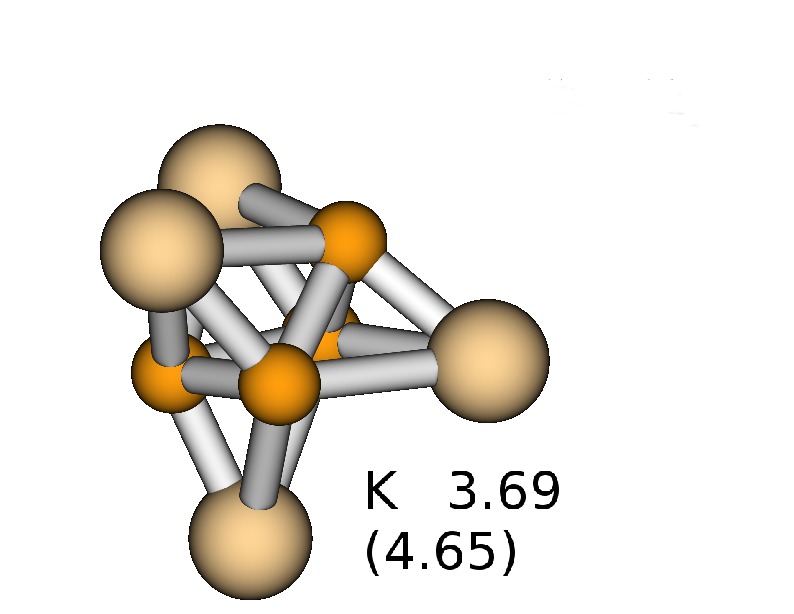}{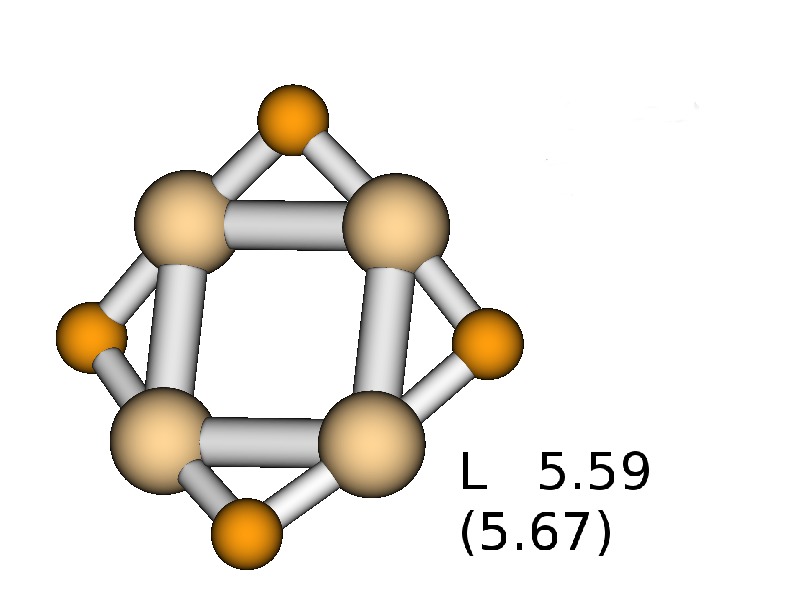}
\plottwo{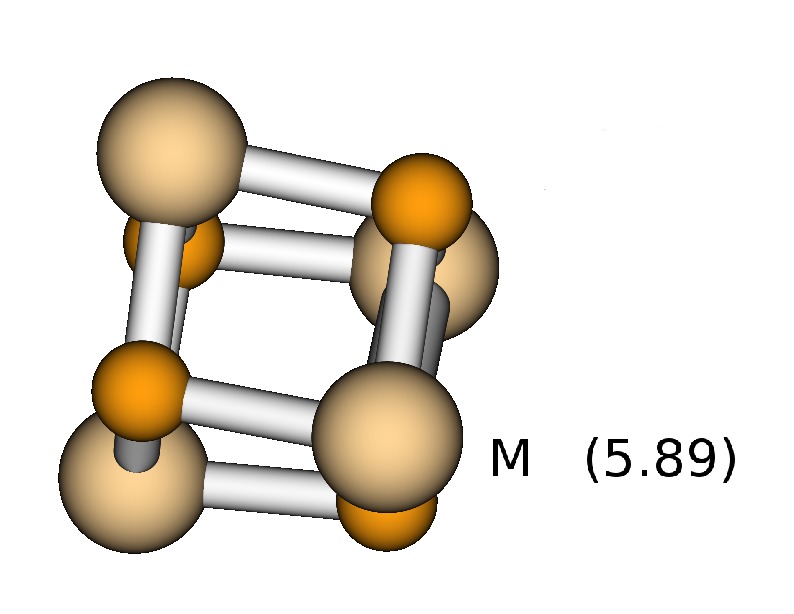}{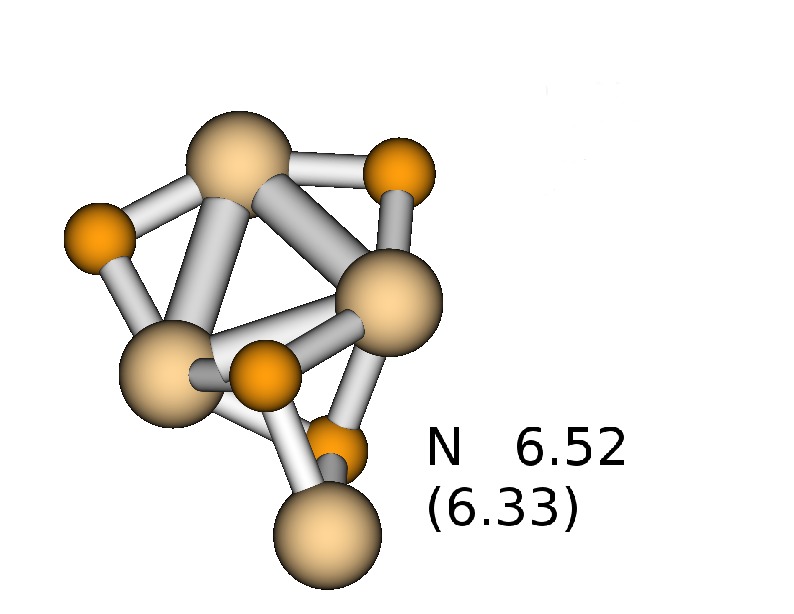}
\caption{Symmetric Si$_4$C$_4$ clusters with alternating Si-C bonds.\label{fig4a}}
\end{figure}

Cluster structures with alternating Si-C bonds have been found by means of Monte-Carlo simulations applying the 
Buckingham pair potential.
These structures show a high degree of symmetry and are displayed in Figure 
\ref{fig4a}. Structure M with T$_d$ symmetry reported by \cite{B902603G} has an energy (5.89) eV above the ground state. The 
other isomers we found have energies ($\sim$ 3-7 eV) far above the ground state. Our results thus indicate that for the size of n=4 the MC-BH generated SiC clusters cannot compete with the segregated structures in Figure \ref{fig4}. An ab-initio exploration of the potential energy surfaces of this cluster size has been performed by \cite{Bertolus2004}.\  

\begin{figure}[h!]
\plottwo{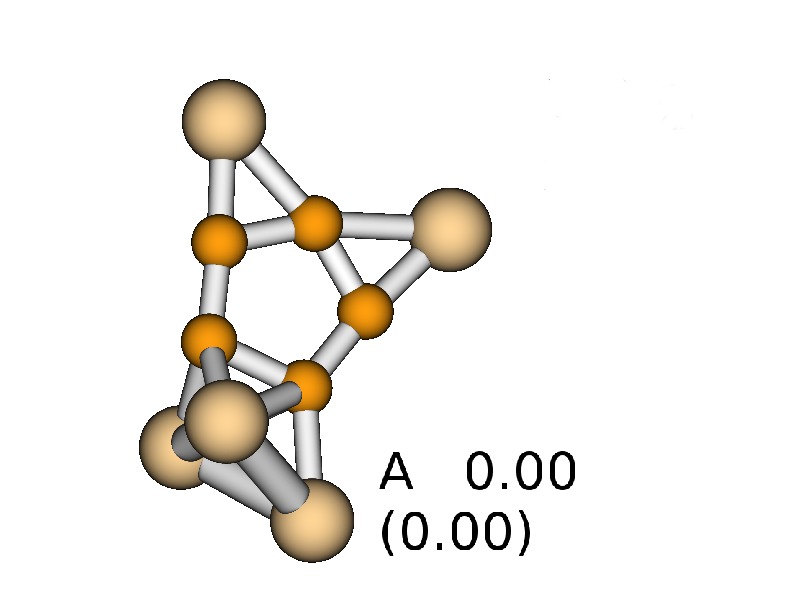}{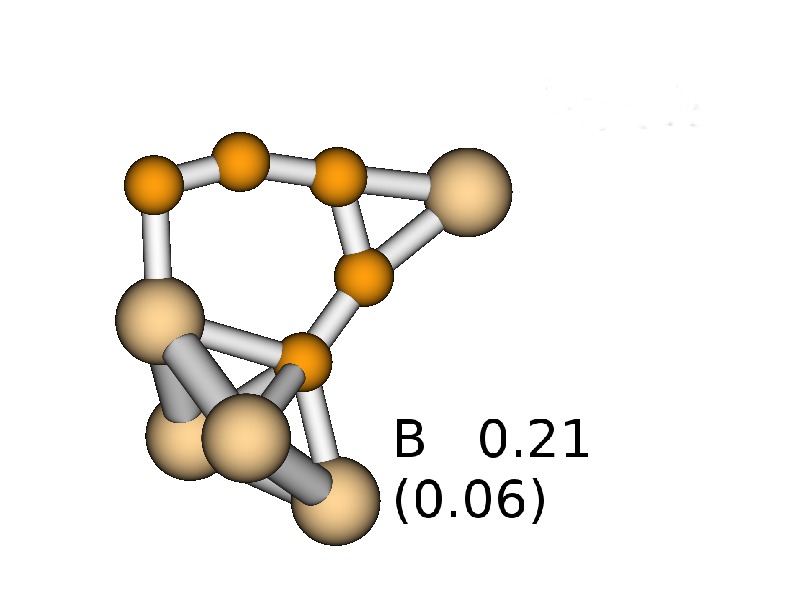}
\plottwo{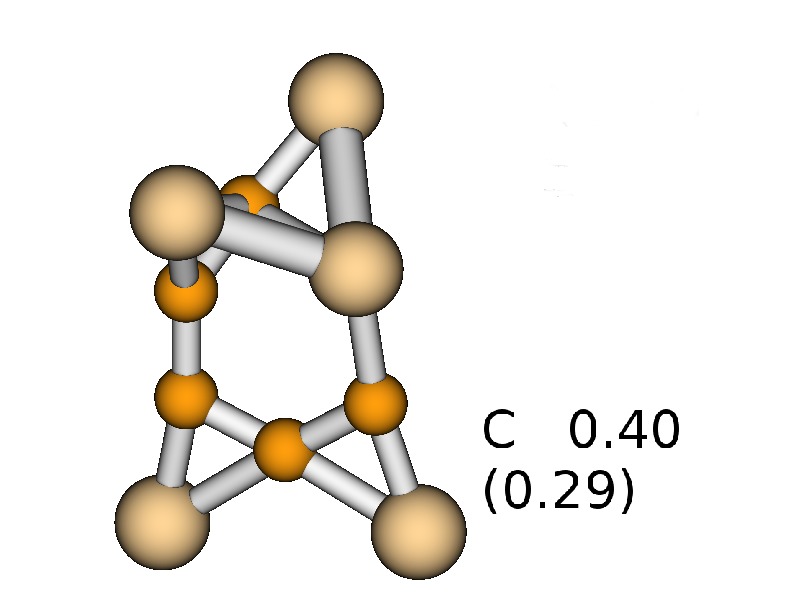}{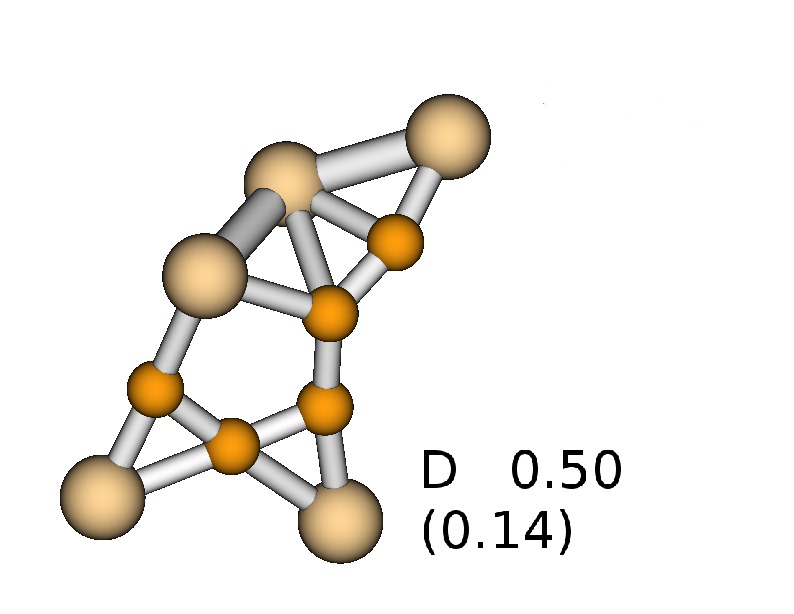} 

\caption{The most stable Si$_5$C$_5$ clusters and relative energies (in eV).\label{fig5}}
\end{figure}

\textbf{Si$_{5}$C$_5$}:\ 
This is the smallest cluster size, where a carbon ring appears. The ground state cluster (A) exhibits a C$_5$ 
ring and a mirror plane, thus belonging to the C$_s$ symmetry group. Structure B shows a C$_s$ symmetry as well, 
and a six member ring with 5 C and 1 Si atoms. Both structures (A and B) have the lowest potential energy in the 
B3LYP calculations as well. 
All low-lying structures displayed in Figure \ref{fig5} exhibit either a bended carbon 
chain with five members or a ring and are non-planar. These structures were previously reported by 
\cite{molecules18078591}. 

\begin{figure}[h!]
\plottwo{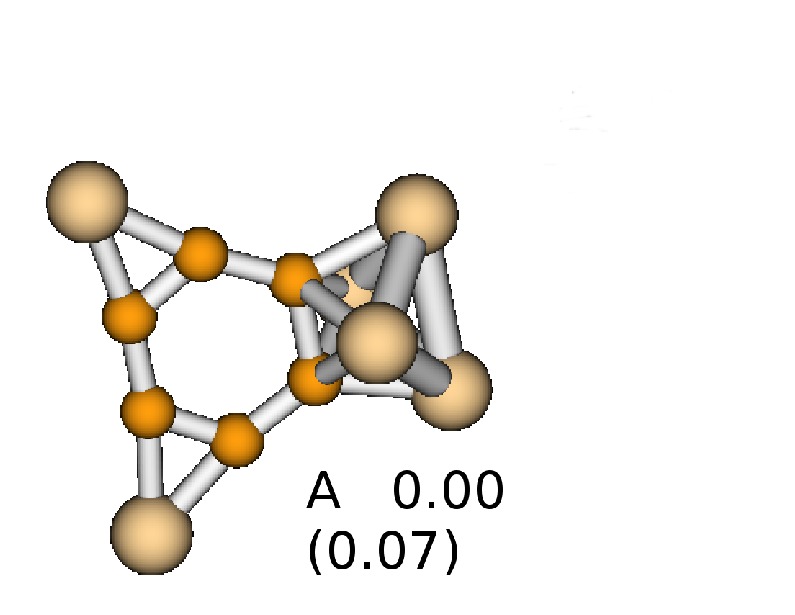}{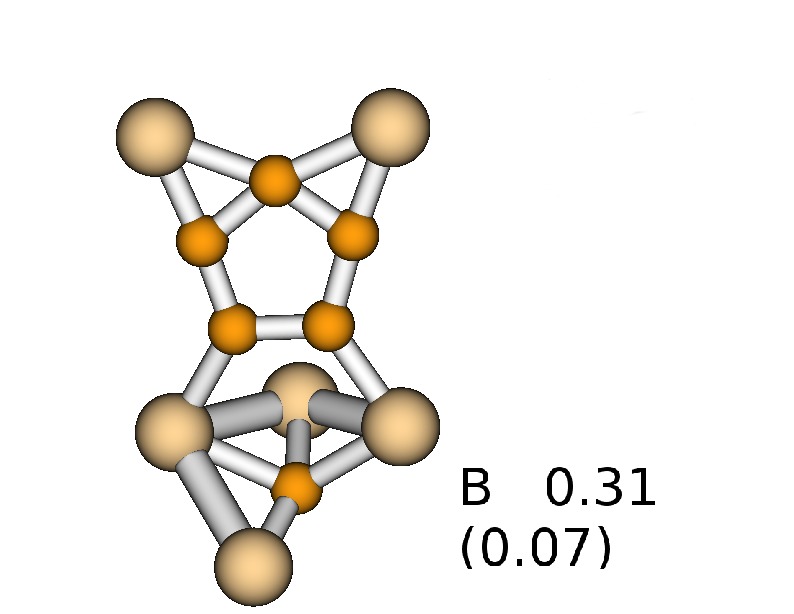}
\plottwo{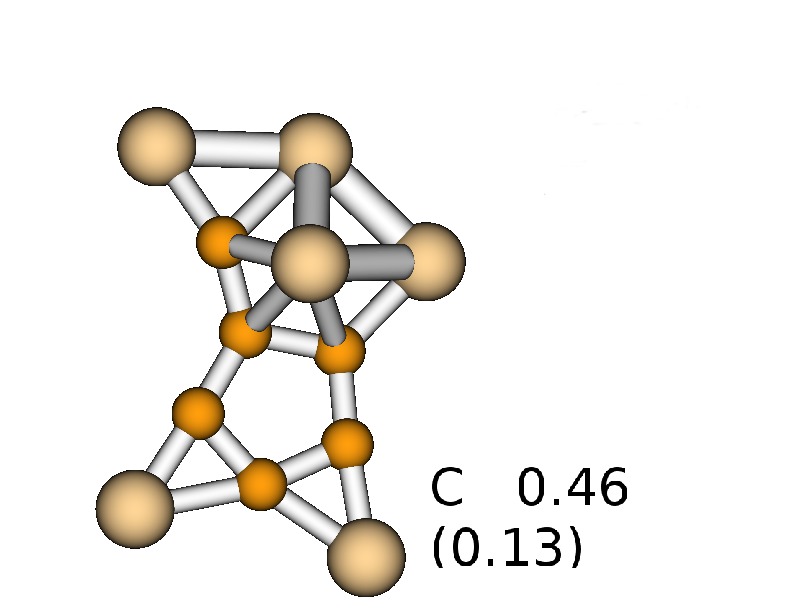}{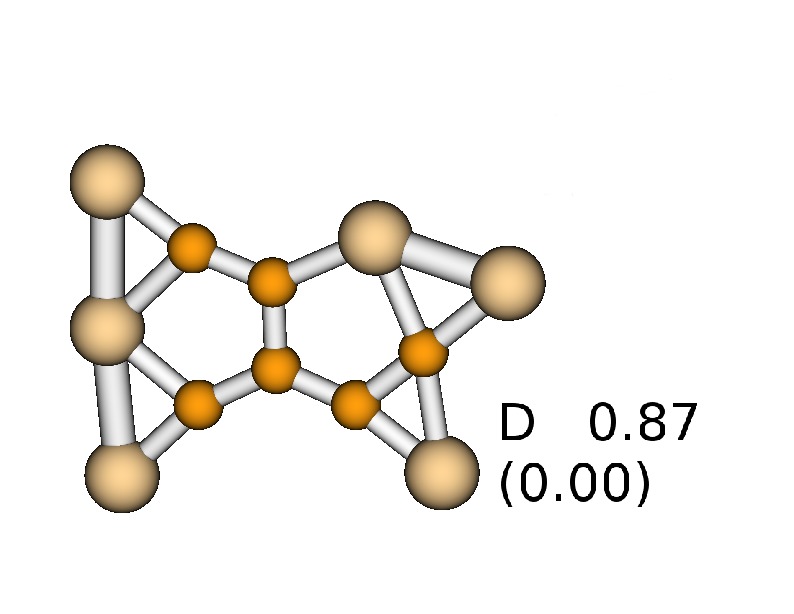}
\caption{The most stable Si$_6$C$_6$ clusters and relative energies (in eV).\label{fig6}}
\end{figure}

\textbf{Si$_{6}$C$_{6}$}:\
The ground state of Si$_6$C$_6$ contains a C$_6$-ring and laterally distributed Si atoms as can 
be seen in Figure \ref{fig6}. Isomer B and and C show 5 member carbon rings.
Structure D has the lowest potential in the B3LYP frame and  is a planar configuration containing two five-
member-rings consisting of four carbon and one silicon atom. Among the lowest lying isomers, compound A 
is the only one found with an aromatic C$_{6}$ ring. The other 
aromatic isomers have significantly higher potential energies. 
We found that structures B and C exhibit a C$_5$ ring and a one-sided silicon segregation (apart from single 
Si atoms conjugating the cluster). 
Clusters obeying a strict alternation of Si and C atoms are 5-9 eV higher in energy compared with the ground 
state. Structures A-D were reported in 
\cite{molecules18078591}.\ 

\begin{figure}[h!]
\plottwo{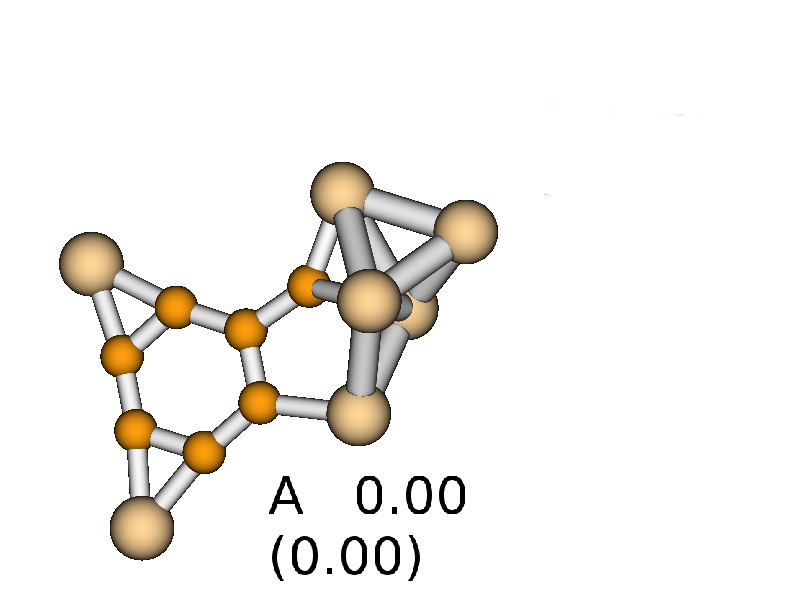}{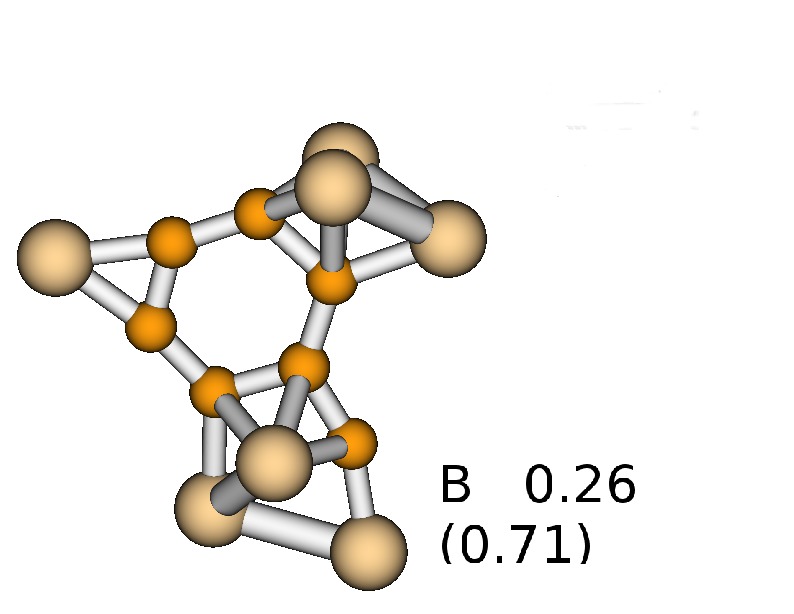}
\plottwo{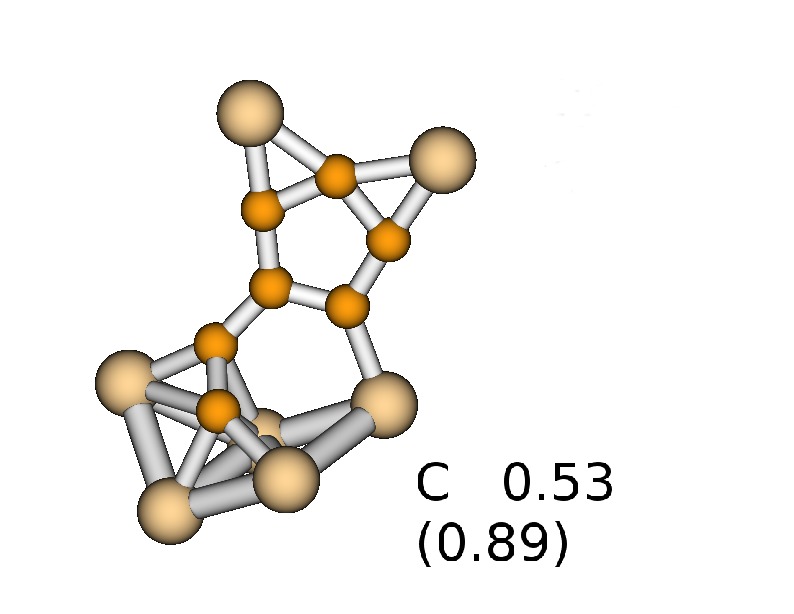}{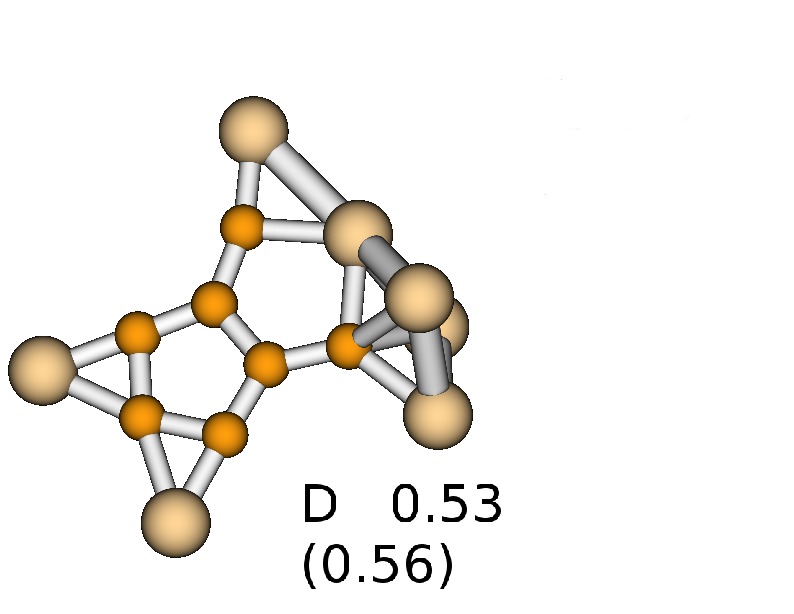}
\caption{The most stable Si$_7$C$_7$ clusters and relative energies (in eV).\label{fig7}}
\end{figure}

\textbf{Si$_{7}$C$_{7}$}:\
The ground  state of Si$_7$C$_7$ (cluster A in Figure \ref{fig7}) consists of aromatic ring connected to a Si$_5$ 
sub-cluster and two individual Si atoms. 
Structures C and D have a five-member carbon ring in common, where the dangling bonds are saturated by two 
individual Si atoms. 
MC-BH generated structures with alternating Si and C atoms have energies 7-8 eV higher than isomer A.\    

\begin{figure}[h!]
\plottwo{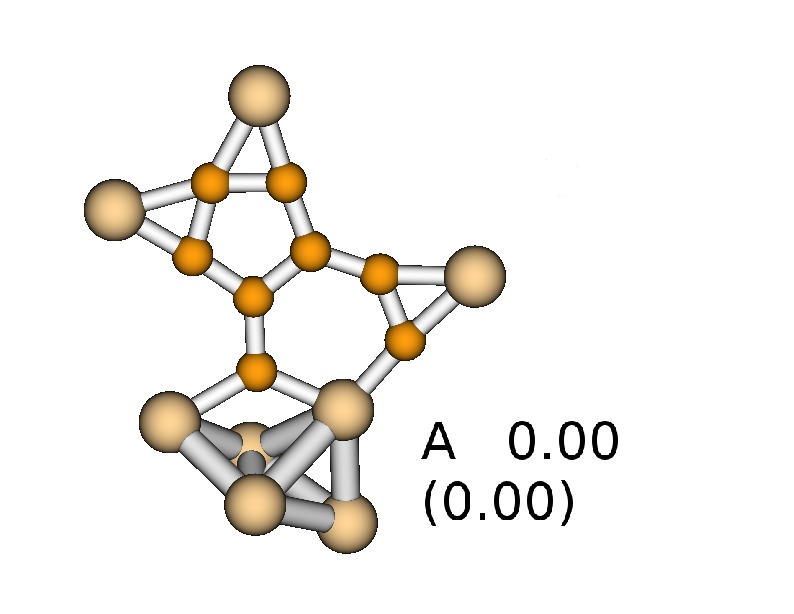}{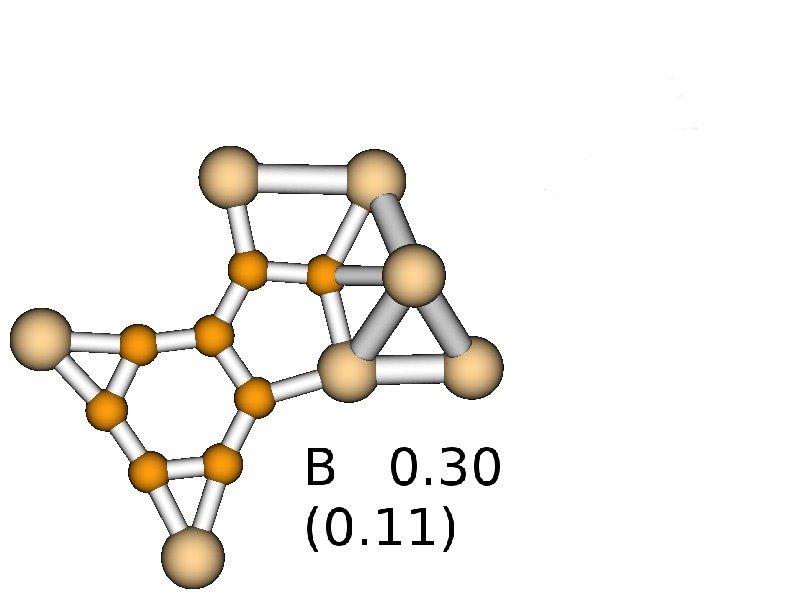}
\plottwo{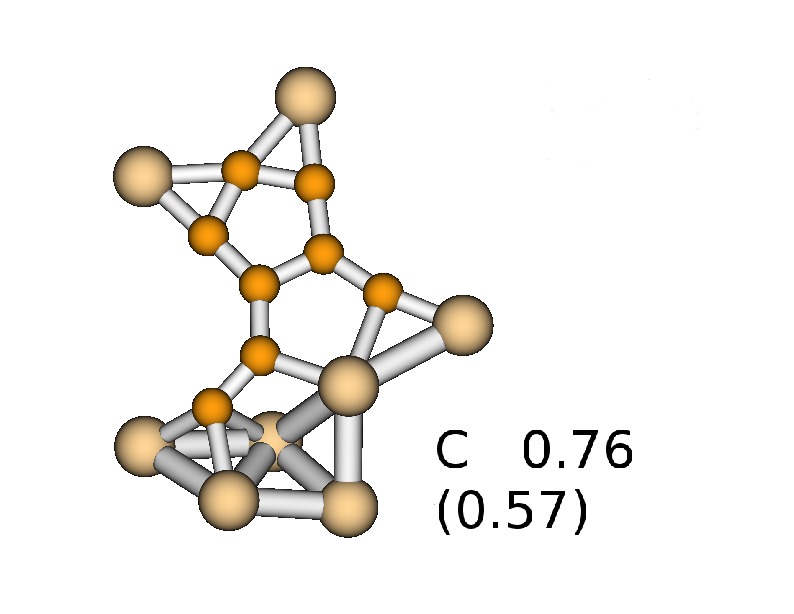}{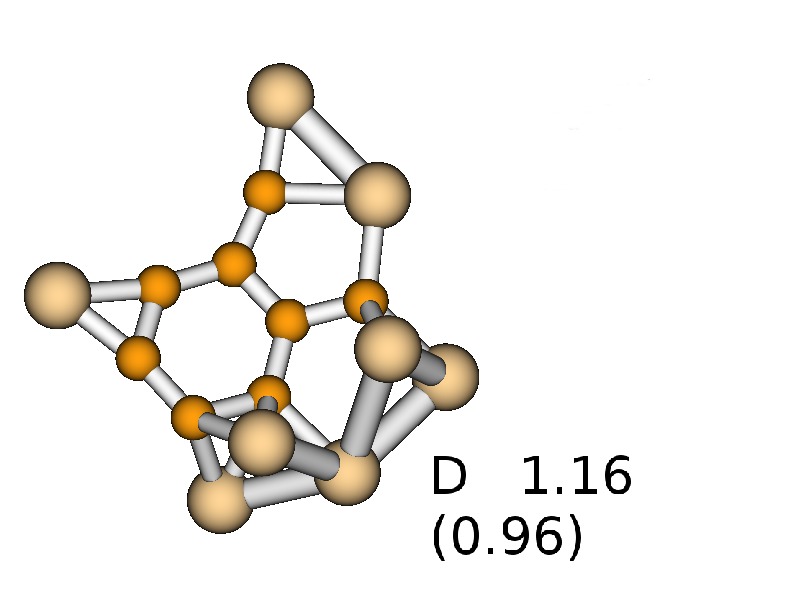}
\caption{The most stable Si$_8$C$_8$ clusters and relative energies (in eV).\label{fig8}}
\end{figure}

\begin{figure}[h!]
\plottwo{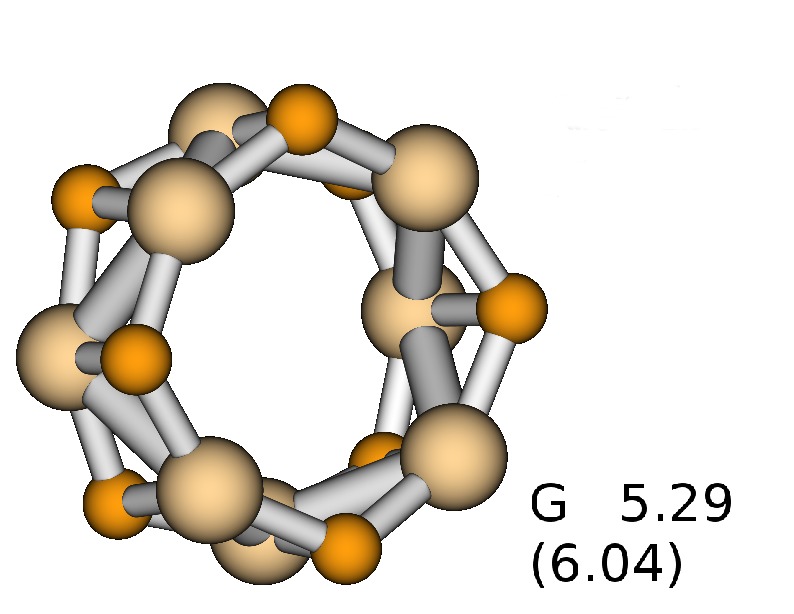}{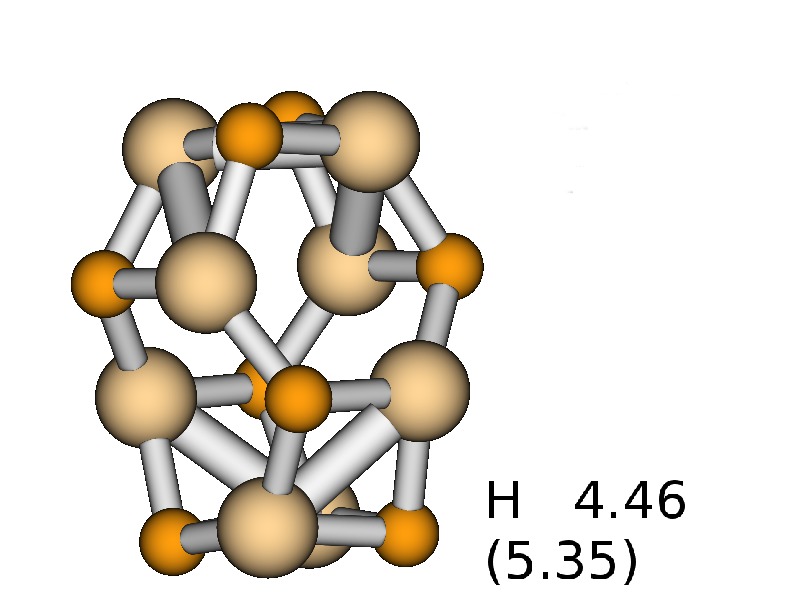}
\caption{Si$_8$C$_8$ cluster structures with altering Si-C bond. Relative energies are given in eV.\label{fig8x}}
\end{figure}

\textbf{Si$_{8}$C$_{8}$}:\
In the most stable Si$_8$C$_8$ clusters the carbon atoms tend to form planar 5- or 6-member rings as can be seen 
in Figure \ref{fig8}. The Si atoms 
surround the carbon sub-cluster and 
segregate spatially. Structures A and C contain a C$_5$-ring, whereas B and D have a C$_6$-ring. The 
remaining carbon atoms arrange as side-chains to form a silicon-substituted ring.     
Structures A,B and C are reported in \cite{molecules18078591}.
Other candidate isomers (except structure D) obtained trough simulated annealing are 3-5 eV above the minimum 
structure.
The highly symmetric double-ring structure (G) proposed by \cite{Belenkov2012} and the ``keyhole'' isomer (H) in
Figure \ref{fig8x}
have potential energies of 5.29 (6.04) eV 
and 4.46 (5.35) eV above the ground state, respectively. Further structures with alternating arrangement of atom types (Si and C) have energies 4-9 eV above the ground state.\

\begin{figure}[h!]
\plottwo{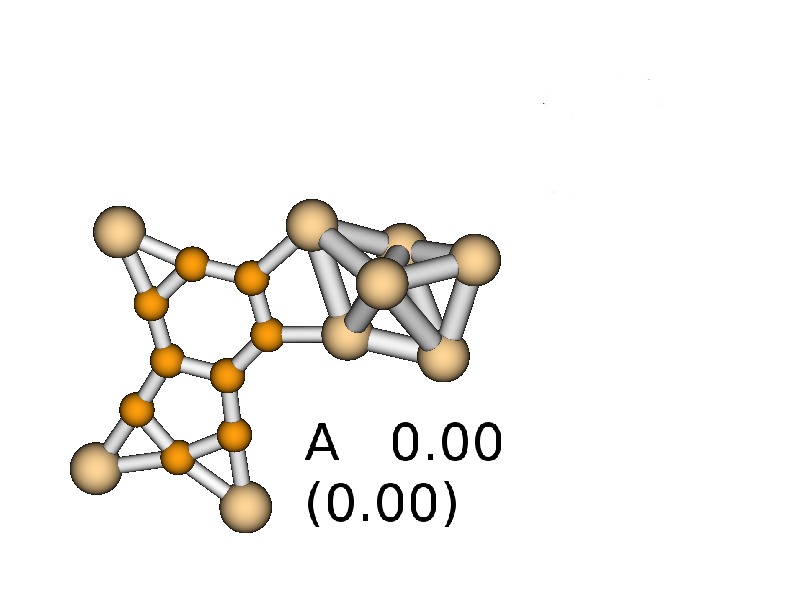}{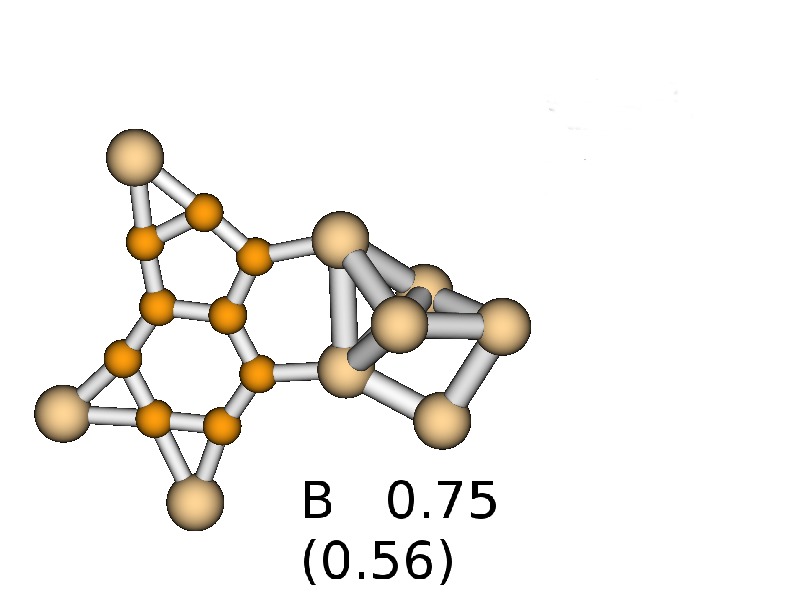}
\plottwo{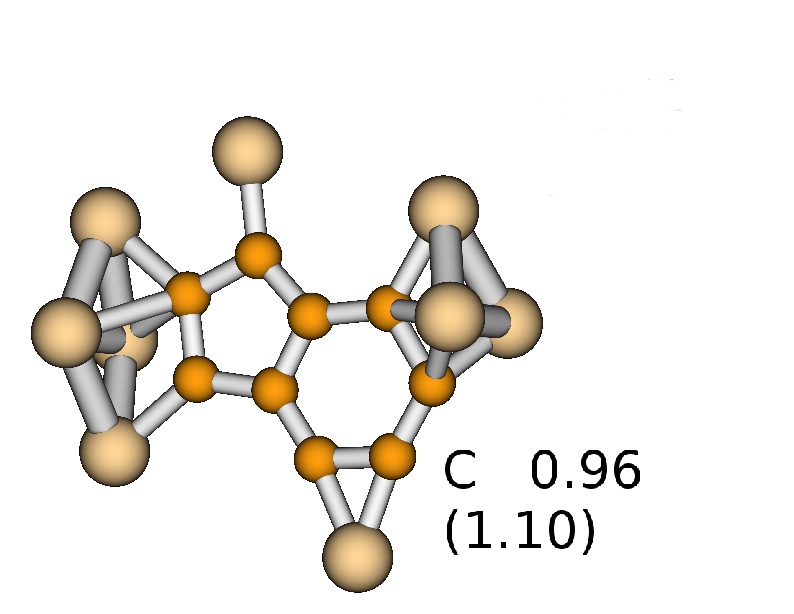}{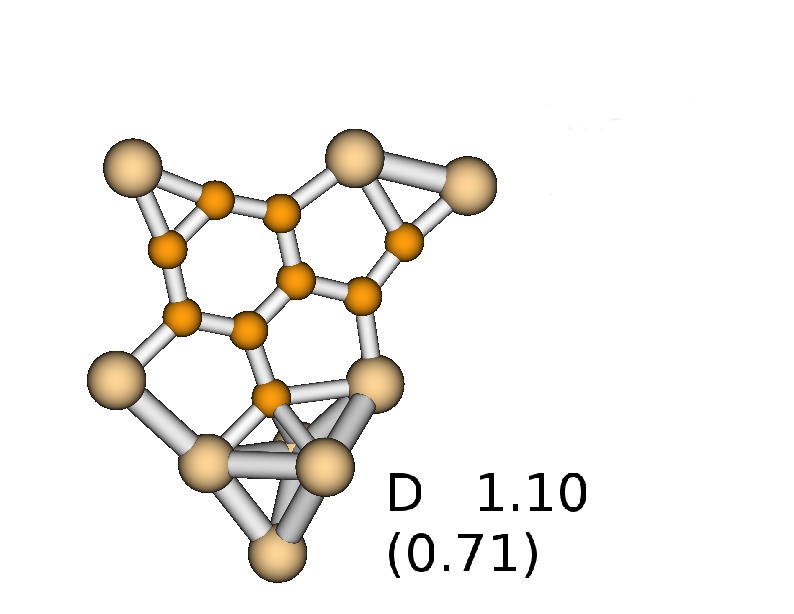}
\caption{The most stable Si$_9$C$_9$ clusters and relative energies (in eV).\label{fig9}}
\end{figure}

\textbf{Si$_{9}$C$_{9}$}:\
The energetically favourable structures A,B and C in Figure \ref{fig9} contain fused C$_6$ and C$_5$ rings. 
Only isomer D has an exceptional character with a C$_6$ and two side chains. 
A C$_{2v}$ symmetric structure with alternating Si and C atoms was obtained by the MC-BH method, but it has an energy of 7.15 (7.84) eV above the ground state. 
Other isomers obtained by simulated annealing have potential energies 2-6 eV above the ground state.\ 

\begin{figure}[h!]
\plottwo{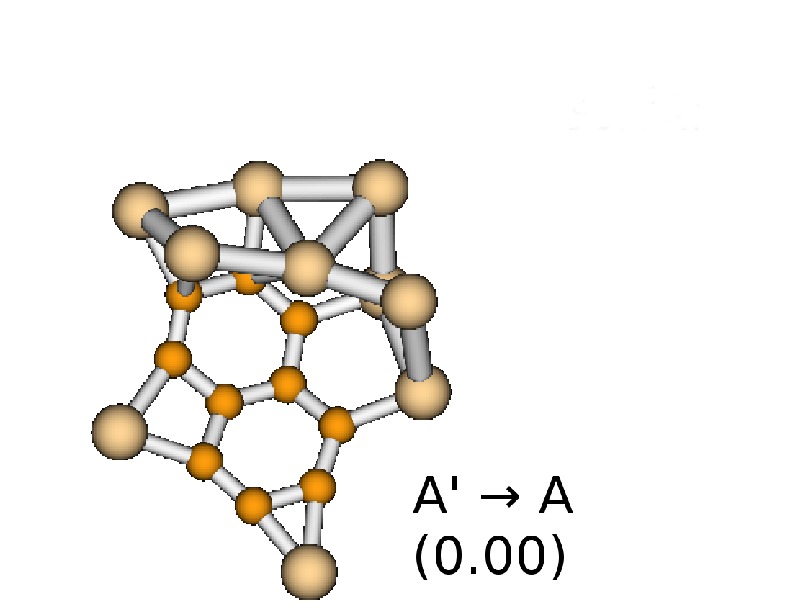}{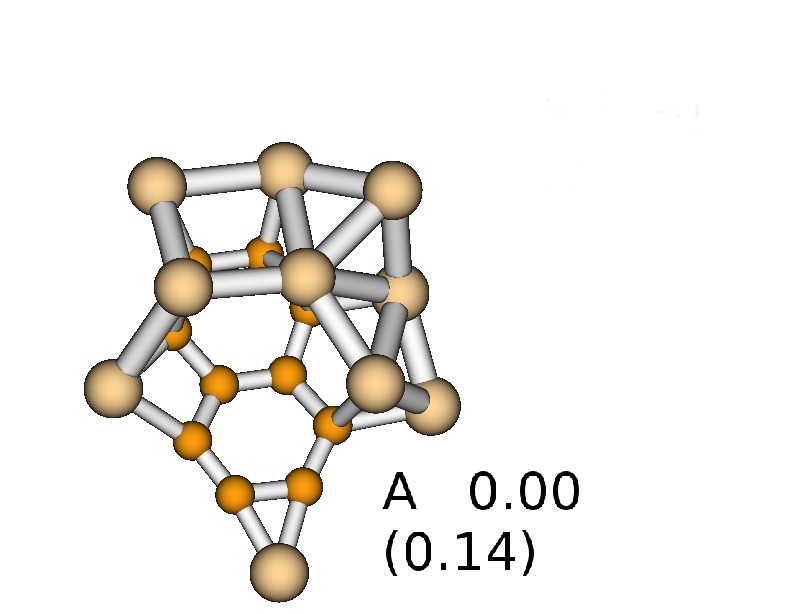}
\plottwo{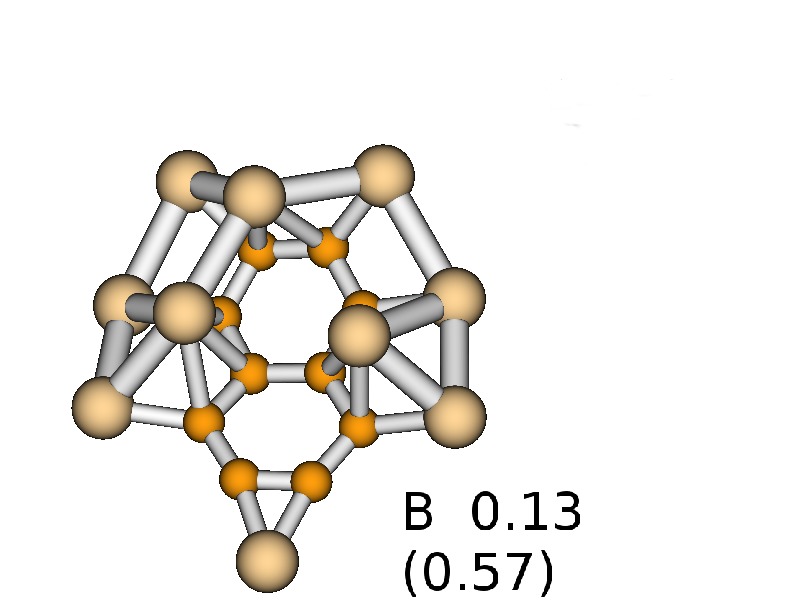}{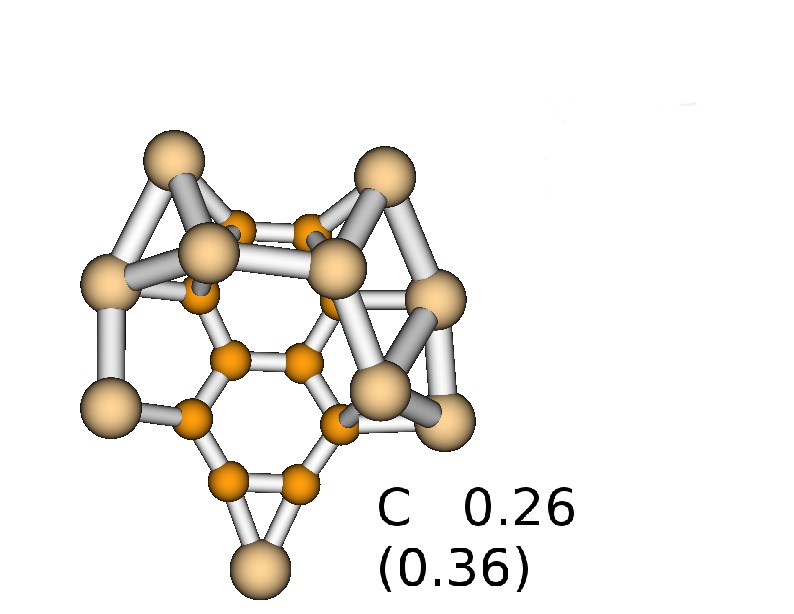}
\caption{The most stable Si$_{10}$C$_{10}$ clusters and relative energies (in eV).\label{fig10}}
\end{figure}

\textbf{Si$_{10}$C$_{10}$}:\
The lowest-lying isomer using the B3LYP functional (structure A') 
is reported for the first time and can be seen in Figure \ref{fig10}. Applying the M11 functional, Structure A' relaxes into 
state A. We thus consider A as the true ground state. Clusters A, B and C
have been found by \cite{molecules18078591}.  
It is prominent that fused double C$_6$ rings of naphthalene type form for the four favourable clusters 
A', A, B, and C.
Compared to smaller SiC cluster sizes, the most stable Si$_{10}$C$_{10}$ clusters have spatial and open cage-like 
forms rather than planar 
configurations. 
We find that the eleven energetically most favourable clusters reside in a narrow energy range of 
1 eV. This is more than for any other size of the investigated SiC clusters. 
Further isomers we found by means of simulated annealing of seed clusters have energies 2-4 eV above the minimum 
energy structure. 
MC-BH synthesised isomers have energies 4-6 eV above the ground state.     

\begin{figure}[h!]
\plottwo{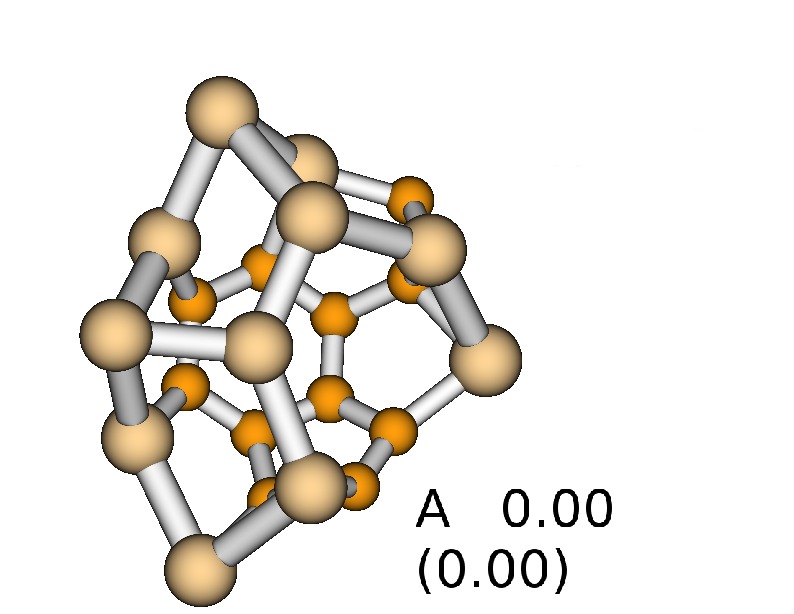}{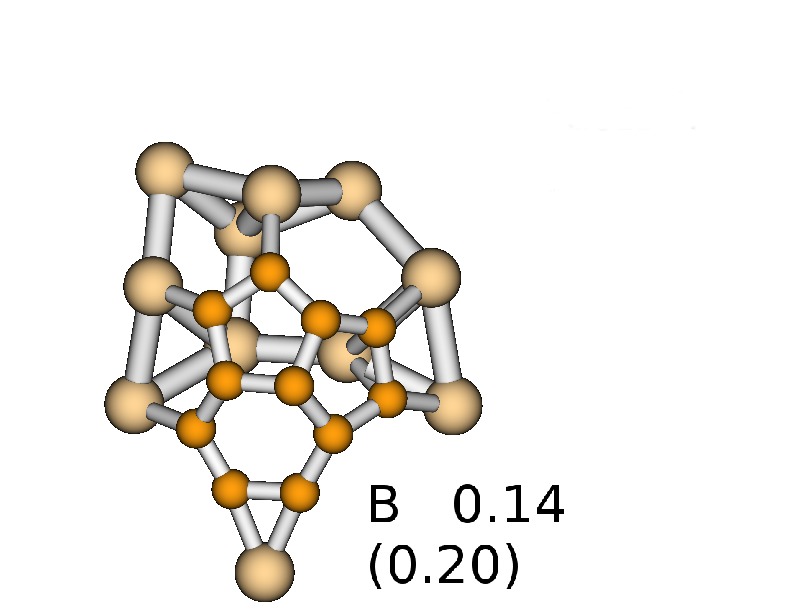}
\plottwo{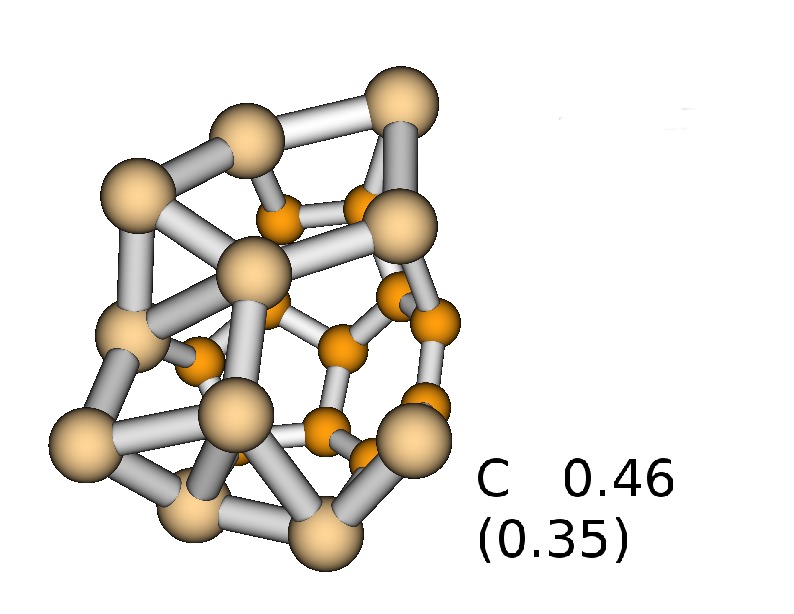}{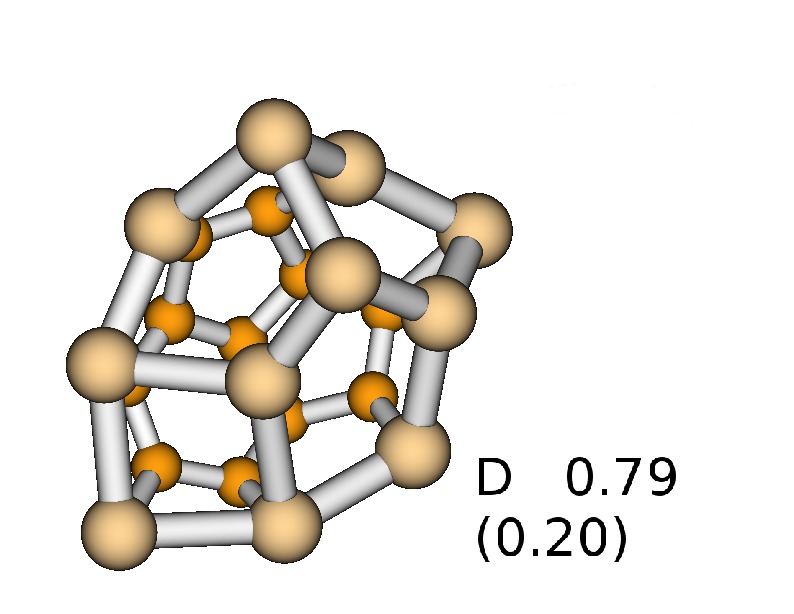}
\caption{The most stable Si$_{11}$C$_{11}$ clusters and relative energies (in eV).\label{fig11}}
\end{figure}

\textbf{Si$_{11}$C$_{11}$}:\
The most stable isomers (see Figure \ref{fig11}) are characterised by presence of a C$_6$ and two C$_5$ rings, respectively, each one sharing an 
edge with another ring. Structures A-D are found in \cite{molecules18078591}. All structures, except the ground state A, have a carbon subunit 
characterised by a fusion of two C$_5$ and one C$_6$ ring. The ground state A contains one  C$_5$ and one C$_6$ ring, and two Si-substituted 
5-member-rings (C$_4$Si rings). 
For this cluster size, stable cage-like clusters found with the MC-BH method possess potential energies 1.2 - 4.5 eV the ground state.\

\begin{figure}[h!]
\plottwo{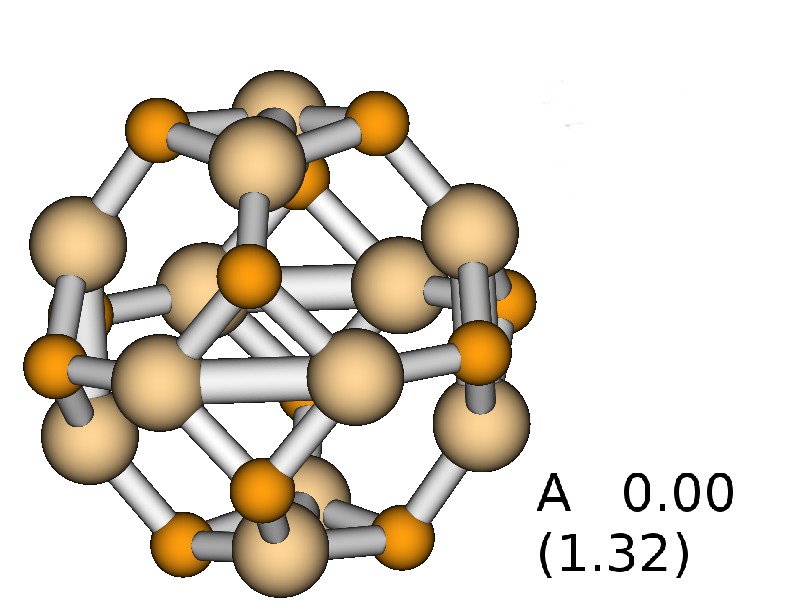}{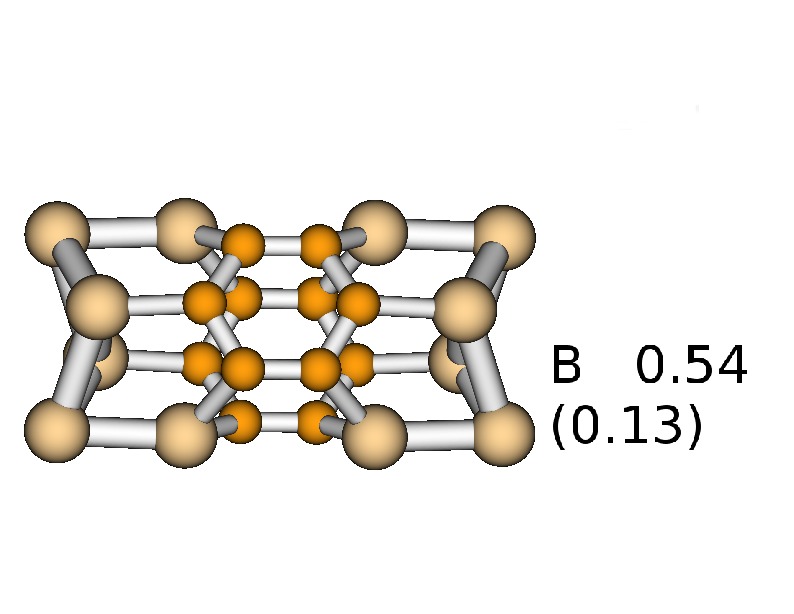}
\plottwo{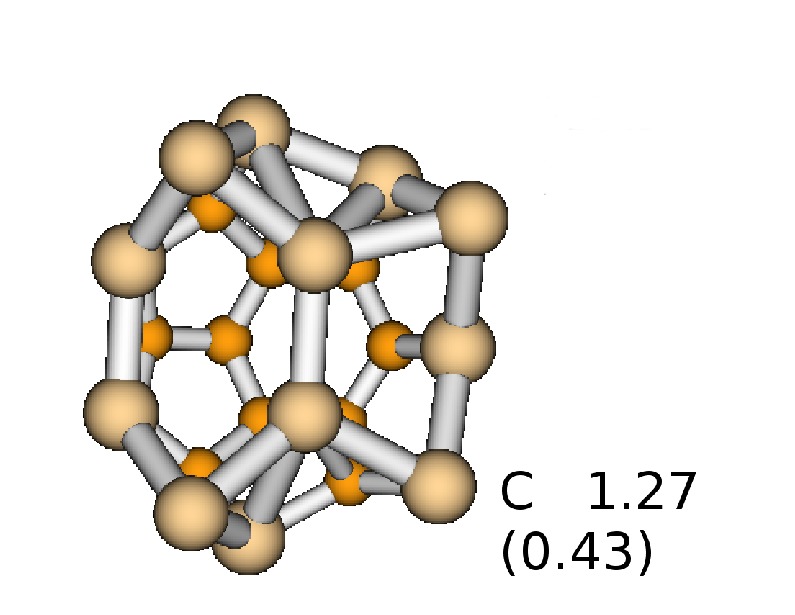}{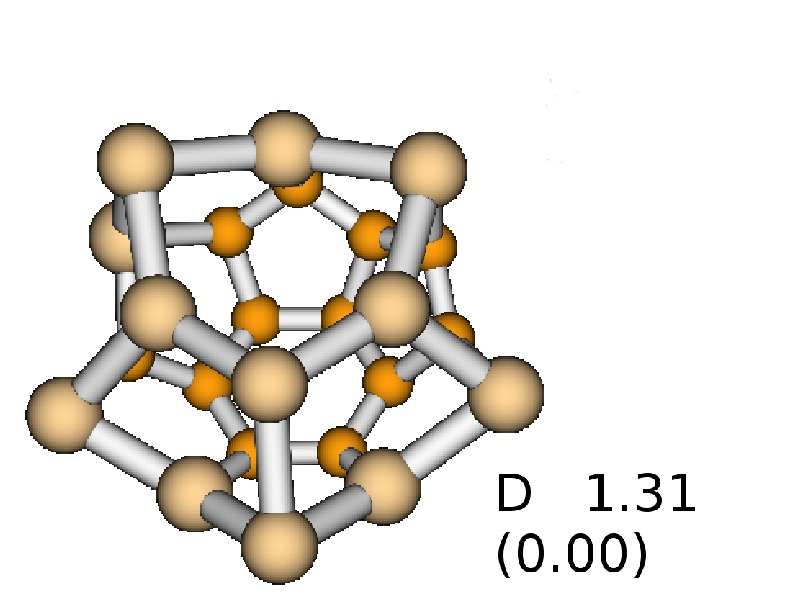}
\caption{The most stable Si$_{12}$C$_{12}$ clusters and relative energies (in eV).\label{fig12}}
\end{figure}

\textbf{Si$_{12}$C$_{12}$}:\ 
The ground state (A) of Si$_{12}$C$_{12}$ is displayed in Figure \ref{fig12} and represents a particular case in 
the series of Si$_n$C$_n$ ground states.
It exhibits a highly symmetric tetrahedral configuration (symmetry group T$_H$) with alternating SiC-bondings and 
was proposed by \cite{B902603G} as a potential nano building block of larger structures. Having a ``bucky''-like configuration, the almost spherical structure resembles the chemical 
family of fullerenes. 
The cluster has a mass of $\sim$ 480 a.m.u. and a diameter of $\sim$ 5.9 \AA . This would result in a mass density of 0.919 g$\cdot$cm$^{-3}$, which is about 30 \% of 3.217 g$\cdot$cm$^{-3}$, a reference mass density for all polytypes \citep{Solr-258823801}.
Owing to its properties, structure A may 
link the segregated clusters with the crystalline bulk material observed in pristine SiC dust grains. 
In contrast to other isomers, structure A may be identified spectroscopically, owing to its strong infrared 
vibration mode intensities (see Figure \ref{figIR}). Moreover, it is the smallest ground state among the most stable Si$_n$C$_n$ 
clusters that we found with the Buckingham pair potential applying the MC-BH method. 
Owing to its stability, shape and atomic coordination, structure A may be a candidate for the basic building blocks of SiC dust grains and may trigger the molecular size where cluster chemistry crosses over to dust chemistry (i.e. condensation and coalescence).

The second most stable polymer (B) exhibits a dihedral D$_{2H}$ symmetric 
structure  with two unconnected C$_6$ rings. 
Despite classified as member of the C$_{1}$ group, cluster D is almost symmetric with a quasi mirror plane. 
These two structures B and D have been reported by \cite{Song2010}. 
Apart from structure C, all these cluster exhibit a fused C$_6$-
C$_6$-C$_5$ ring segregation. 
Isomers C has not been reported previously.

We found several other stable clusters with higher potential energies. However, they 
are not displayed as they have a similar open-cage like configurations such as structures C and higher lying isomers. 	
Among the lowest lying Si$_{12}$C$_{12}$ configurations a high degree of 
symmetry is prominent.
The symmetric structures obtained with the MC-BH method are, apart from the ground state, 8-9 eV higher in 
potential energy than isomer A.\ 

\begin{figure}[h!]
\plottwo{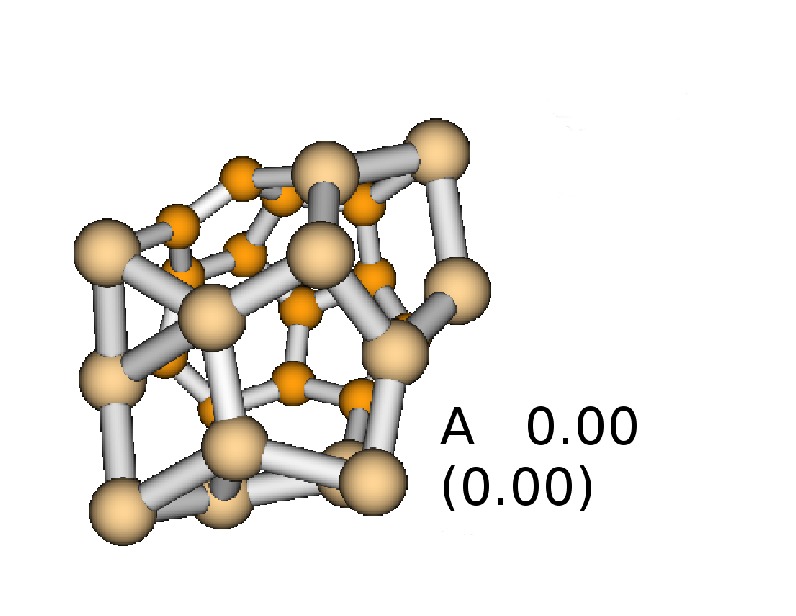}{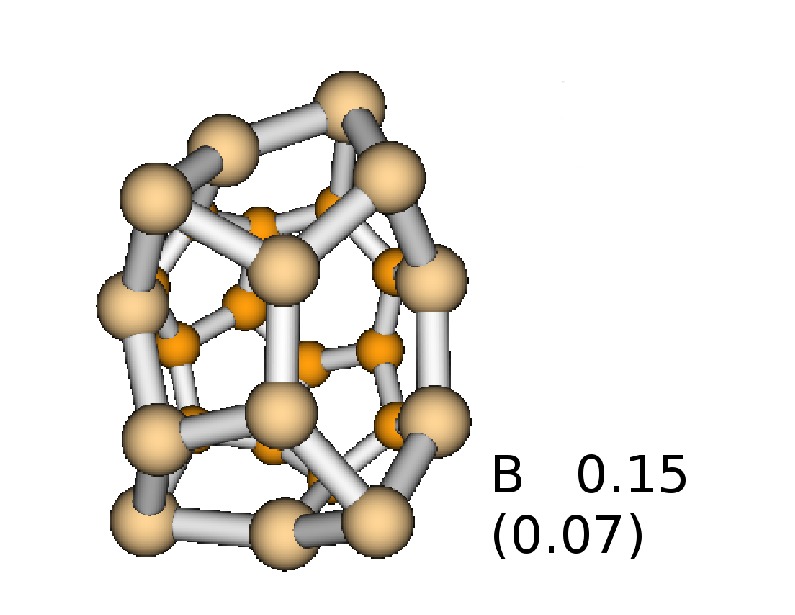}
\plottwo{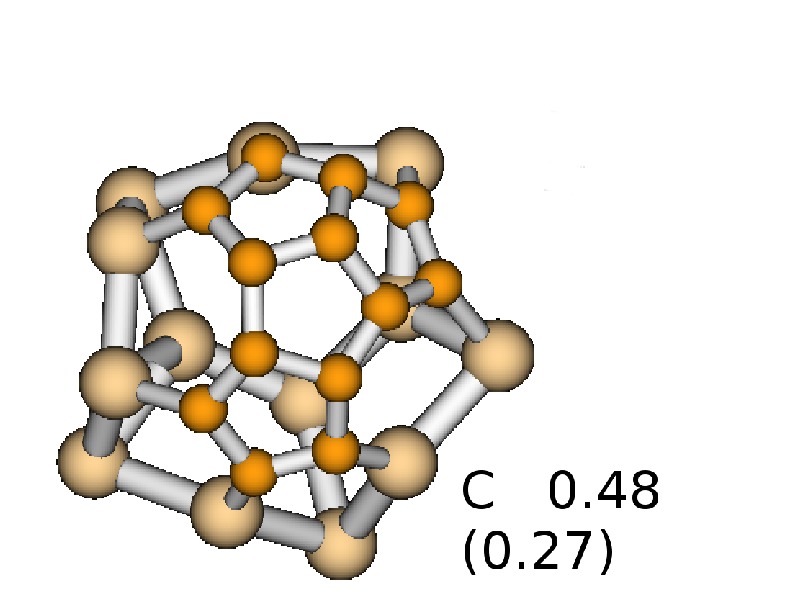}{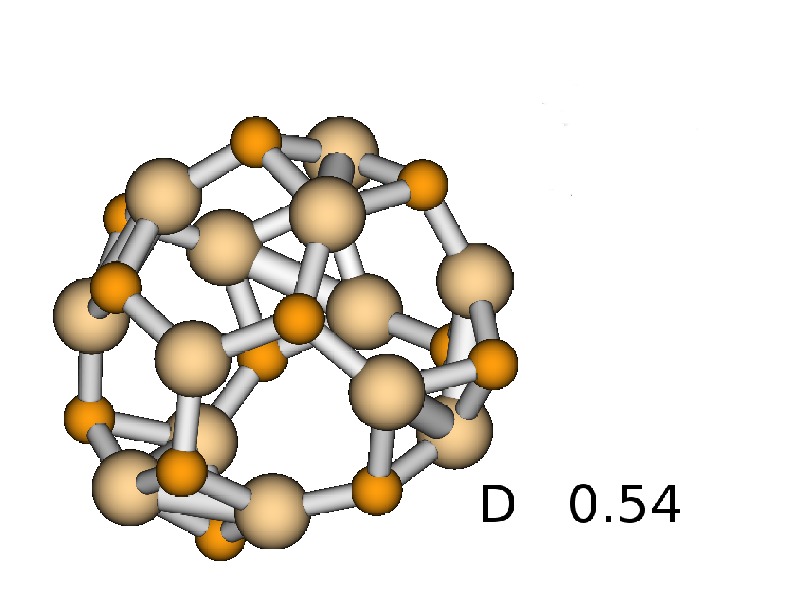}
\caption{The most stable Si$_{13}$C$_{13}$ clusters and relative energies (in eV).\label{fig13}}
\end{figure}

\textbf{Si$_{13}$C$_{13}$}:\ 
Among the lowest-lying Si$_{13}$C$_{13}$ isomers, we find carbon segregations with one C$_6$ and three C$_5$ 
rings (structures A and B) and 
four C$_5$ rings (C) (see Figure \ref{fig13}). Also the Si atoms start to develop segregated rings with 5 to 6 
members (A,B and C). 
A, B and C have been found by \cite{Song2010}, however, owing to the different functional/basis set used in their 
study, in a different energy ordering. We find that the energetic ordering is preserved by comparing our M11 and 
B3LYP results. Further structures found by simulated annealing are 0.5-4 eV higher in potential energy
Structure D is a MC-BH generated structure and contains seven 6-member rings and 1 large 8-member ring. 
We find three further structures with cage-like geometries that have potential energies 0.56-1.93 eV 
above the ground state and thus, they are comparable to the segregated clusters obtained by simulated annealing.\

\begin{figure}[h!]
\plottwo{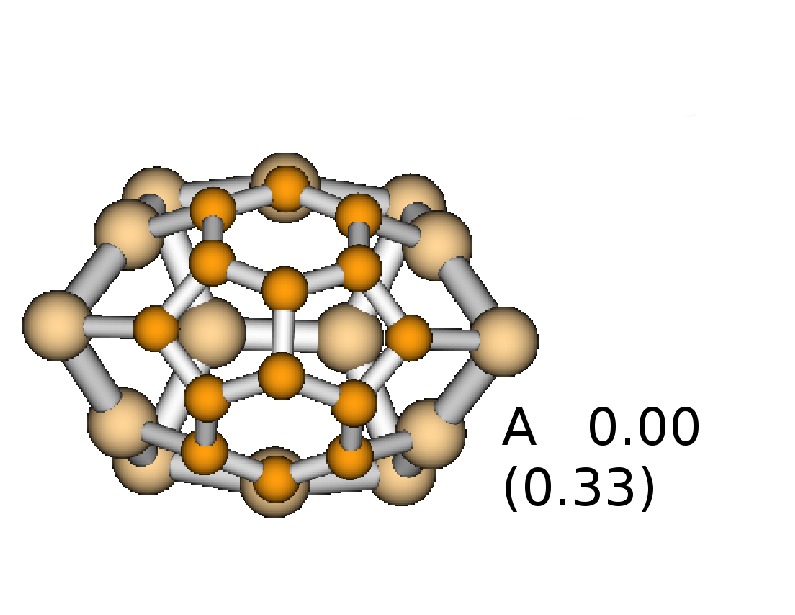}{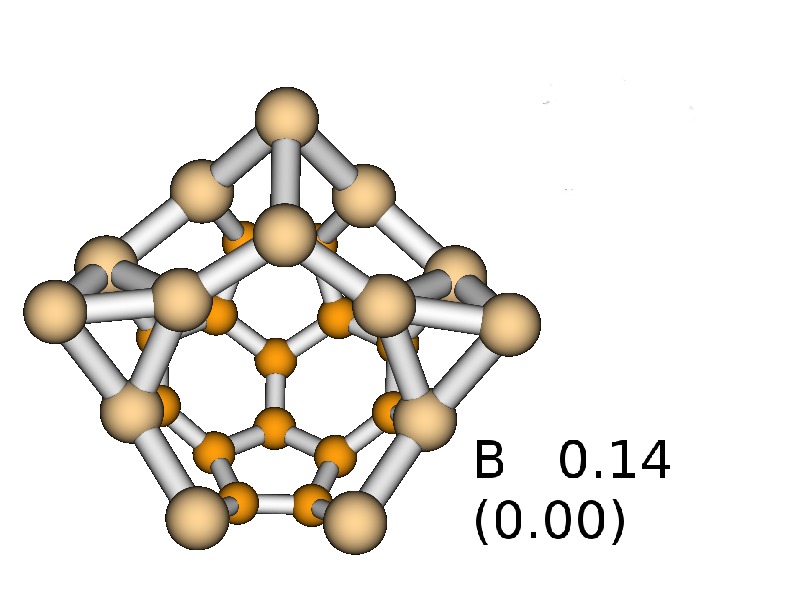}
\plottwo{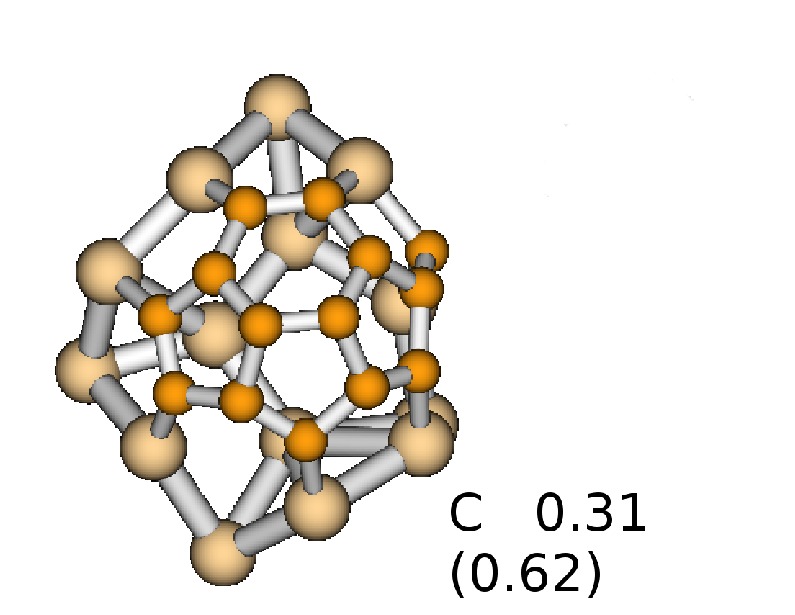}{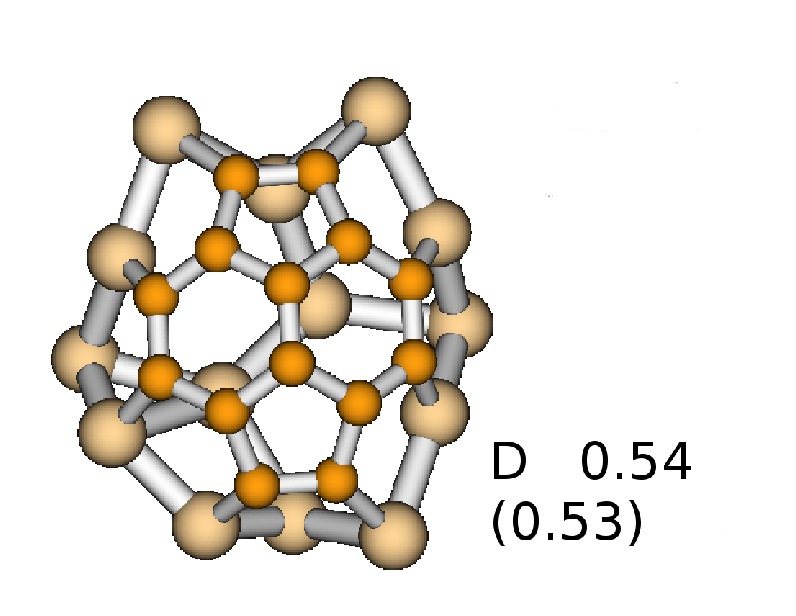}

\caption{The most stable Si$_{14}$C$_{14}$ clusters and relative energies (in eV).\label{fig14}}
\end{figure}

\textbf{Si$_{14}$C$_{14}$}:\ 
The two most stable isomers of Si$_{14}$C$_{14}$ both show symmetries as can be seen in Figure \ref{fig14}. 
The first (A) shows two mirror 
planes and a two fold symmetry axis (C$_{2v}$ group), the second (B) has one mirror plane (C$_s$ group). The  
most stable structures (A, B and D) show a complete carbon segregation consisting of 
two C$_6$ and two C$_5$ rings. Whereas in B and D the C$_6$ rings are connected and share a C-C bonding, in A the 
C$_5$ rings share binding 
electrons and the C$_6$ rings are separated from each other. Moreover, B and C show an overall open cage 
geometry, whereas A represents a closed 
hollow ellipsoid. Isomers A, B and C were found in \cite{Song2010}, D is reported for the first time. 
Further structures obtained by simulated annealing are not displayed here and have 1-4 eV above the ground state.
With the MC-BH approach we found a structure with alternating Si-C bonds and an energy 0.64 eV above the ground state. 
It is the fifth lowest energy structure for Si$_{14}$C$_{14}$. 
Other stuctures obtained with the MC-BH method have energies 1.4 - 1.7 eV above the global minimum. The latter findings
indicate that for n=14 the segregated forms cease to dominate and cage-like clusters can compete against the latter in 
terms of potential energy .\

\textbf{Si$_{15}$C$_{15}$}:\ 
The ground state of Si$_{15}$C$_{15}$ is a symmetric structure with alternative bond Si-C bonds found with the MC-BH method. 
It is composed of eleven 6-member  and four 4-member rings obeying strict alternation of Si and C atoms. This ground state is reported for the first time. Its low potential energy indicates that the symmetric cage-like configurations with alternating Si-C bonds can compete against segregated structrures at this size regime and even represent the lowest energy structure for n=15. 

\begin{figure}[h!]
\plottwo{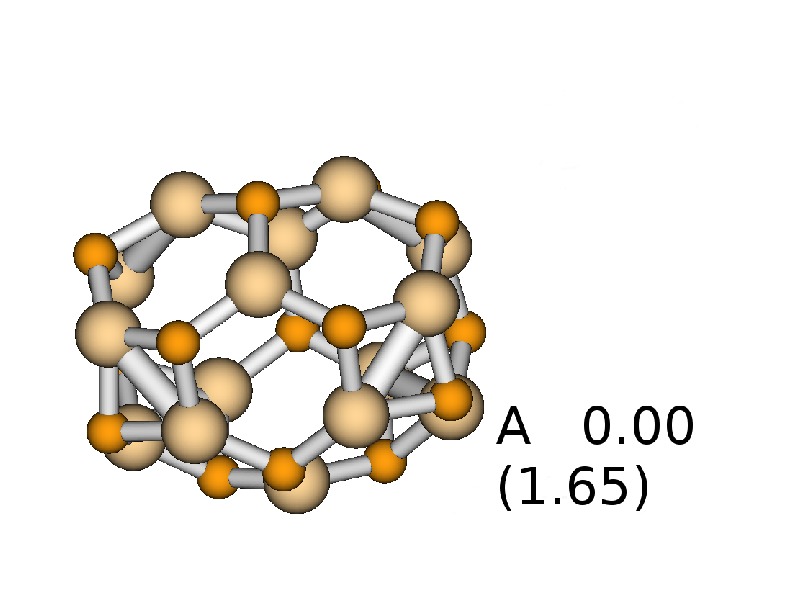}{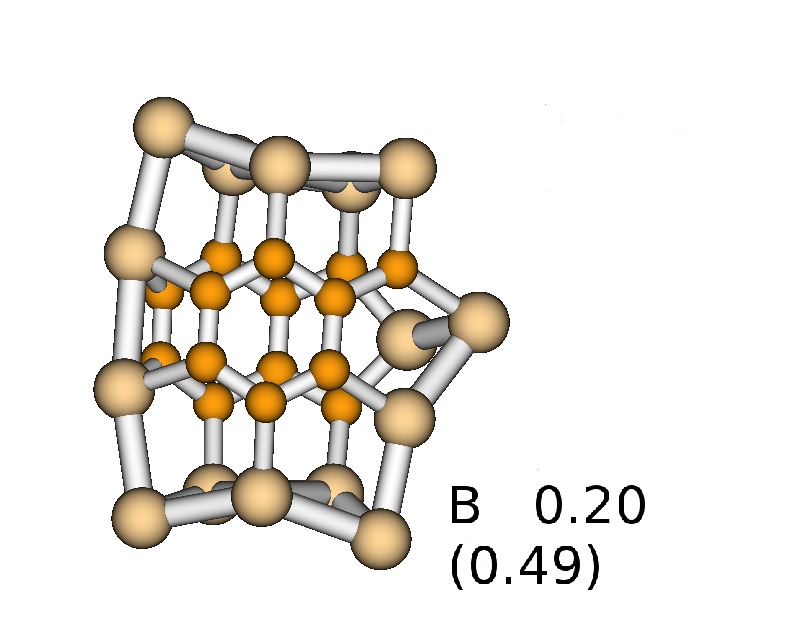}
\plottwo{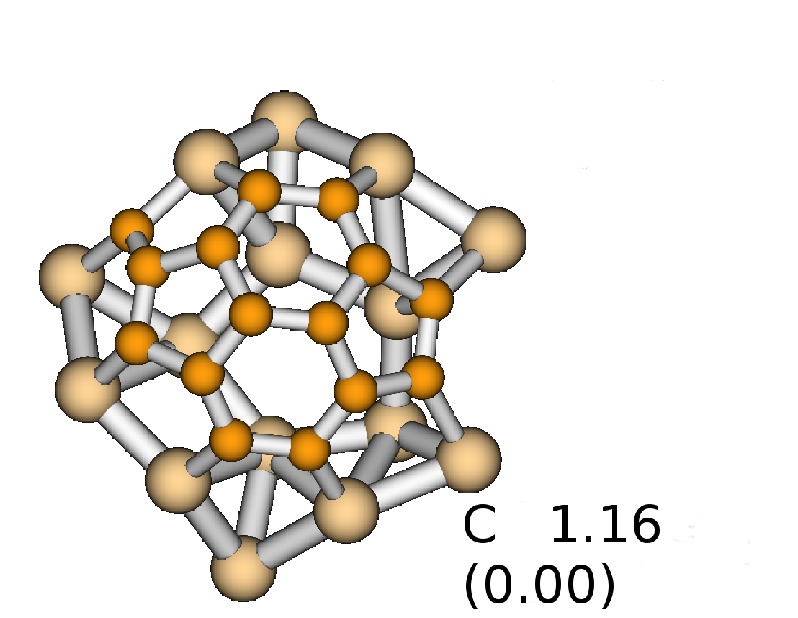}{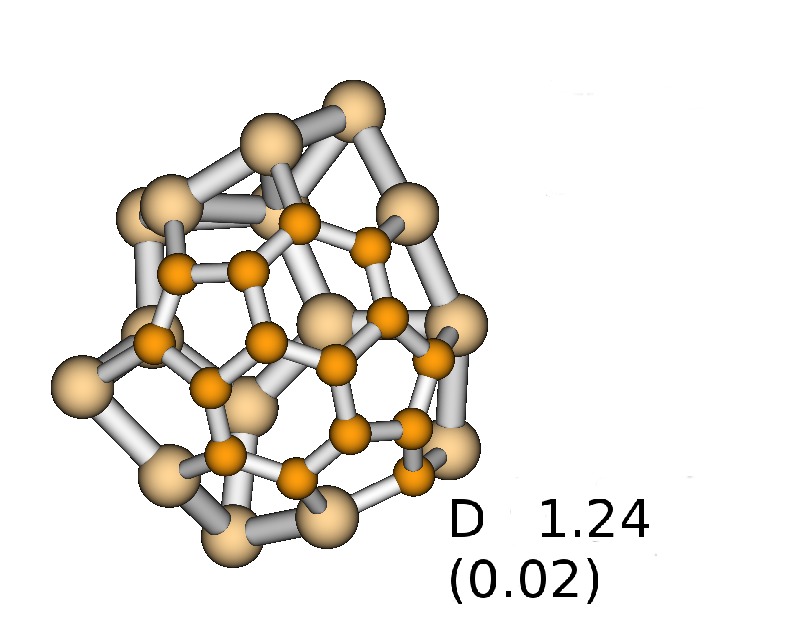}
\caption{The most stable Si$_{15}$C$_{15}$ clusters and relative energies (in eV).\label{fig15}}
\end{figure}

The second lowest isomer showing two separate carbon segregations and a quasi-mirror plane is displayed in Figure \ref{fig15}. It shows similarity with the smaller-sized (n=12) structure B in Figure \ref{fig12}.
The two low-lying isomers (C and D) are almost identical in terms of their B3LYP potential energy and can be termed 
degenerate. The investigation of these cluster with the M11 functional, however, reveals a larger spacing in energy, and that they 
can be regarded as independent and discrete clusters.  
Moreover, they show distinct geometries, vibrational IR spectra and rotational constants.  
Their carbon subunit is almost identical and resembles structure D in Figure 6 of \cite{Song2010}. 
However, the silicon atoms are arranged 
differently, giving rise to the change in potential energy.  

Structure B has been reported previously by \cite{Song2010}.\

\textbf{Si$_{16}$C$_{16}$}:\
The ground state of Si$_{16}$C$_{16}$ shows a particular high degree of symmetry (point group T$_d$) and is composed of 6-member rings with alternating Si-C bondings. The structure has been put forward as building units of larger SiC frameworks by \cite{B902603G}.
The overall structure is a closed hollow fullerene-like cage and shows strong IR features (around 9.3-9.5 and 18.9 $\mu$m, see Figure \ref{figIR}), compared to isomer B (and the other isomers of this size). As for n=12, we found the ground state structure by applying the Buckingham pair potential using MC-BH.
The ``bucky''-like structure has an approximately spherical shape and exhibits alternating Si-C 
bonds.
The hollow spheres with a T$_d$ symmetry have a mass of $\sim$ 640 a.m.u. and a diameter of $\sim$ 6.5 \AA . This would result in a mass density of 0.924 g$\cdot$cm$^{-3}$, which is very similar as for the n=12 case and about 30 \% of 3.217 g$\cdot$cm$^{-3}$, a reference mass density for all polytypes \citep{Solr-258823801}.
Owing to its strong and characteristic infra-red features, this particular isomer can be spectroscopically identified.
As for n=12, the ground state (A) may link the segregation-dominated small clusters (n$<$12) with larger clusters and crystalline SiC bulk material.  

\begin{figure}[h!]
\plottwo{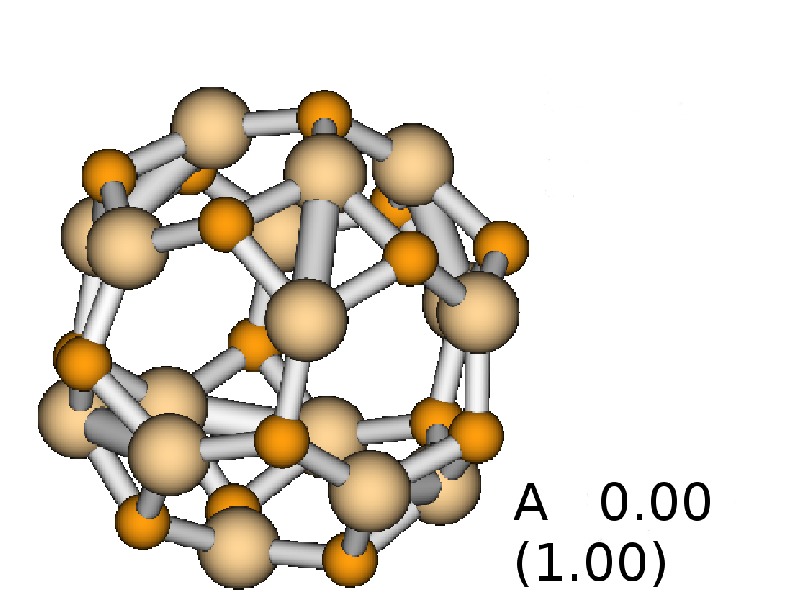}{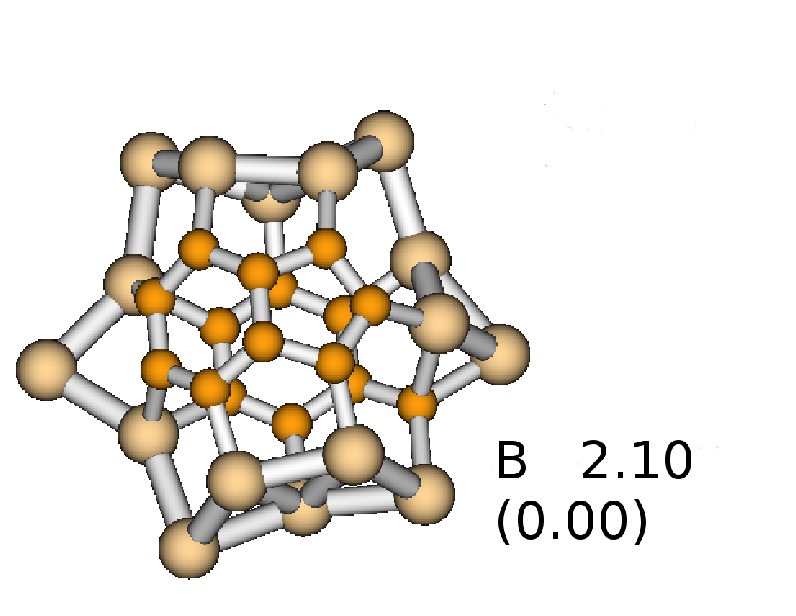}
\plottwo{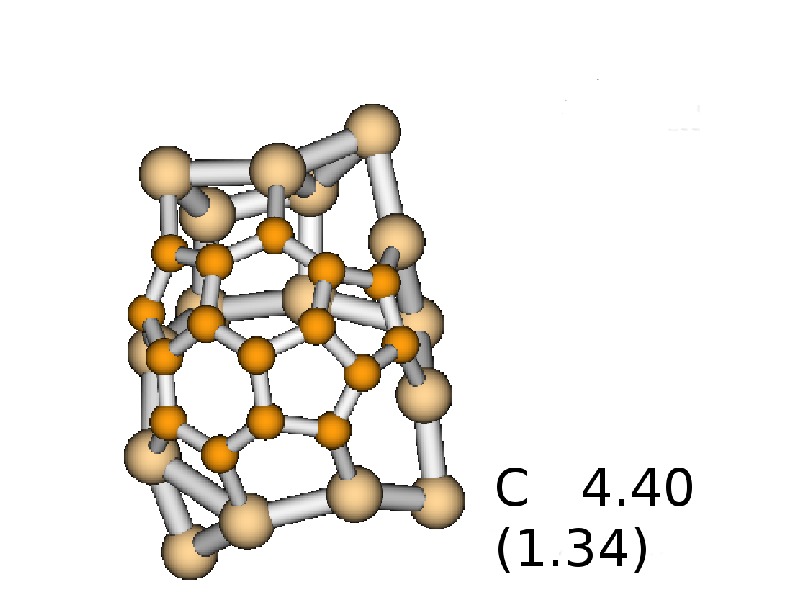}{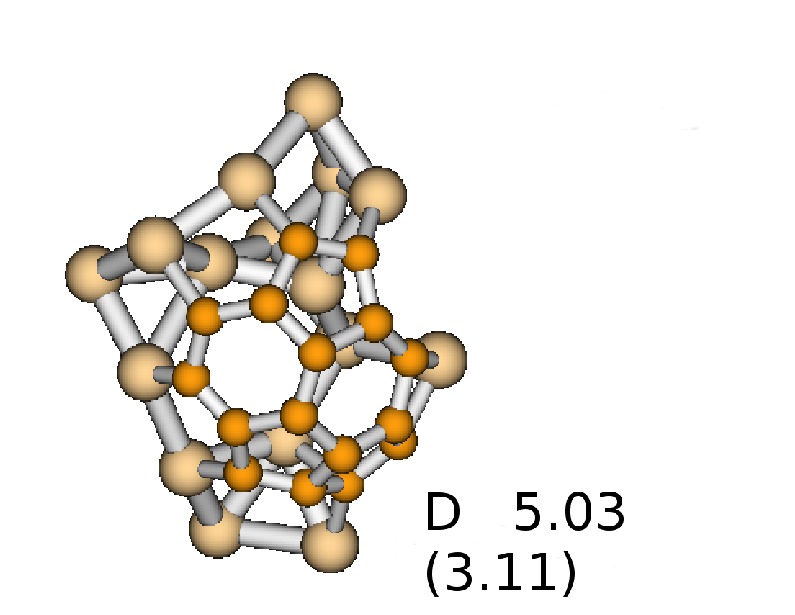}
\caption{Stable isomers of Si$_{16}$C$_{16}$ clusters.\label{fig16}}
\end{figure}

\begin{figure}[h!]
\plotone{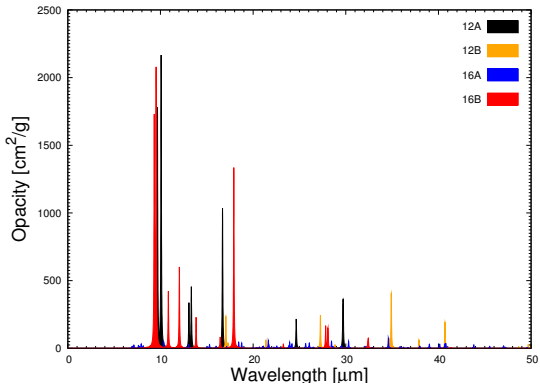}
\caption{Vibrational infrared spectra of the two lowest-lying states of Si$_{12}$C$_{12}$ and Si$_{16}$C$_{16}$. 
12A and 16B correspond to the ``bucky''-like configurations. 
\label{figIR}}
\end{figure}

The next higher lying isomer of Si$_{16}$C$_{16}$ (structure B) shows two carbon segregations, a C$_6$C$_5$-ring and a C$_6$-ring with a one-C-atom arm, and 
exhibits a distorted symmetry with a quasi mirror plane, as can be seen in Figure \ref{fig16}. 
Isomer B has been found for the first time and is the lowest energy structure using the B3LYP functional. 
Despite its low potential energy, it is challenging to observationally detect structure A, owing to low vibrational IR 
intensities (see Figure \ref{figIR}).

For sizes n$\le$12 we clearly show the emergence of a new family of stable clusters, the cage-like structures with alternating Si-C bonds.
Some properties (alternating atomic arrangement, bond lengths) of the latter strongly resemble the bulk phase of 3C-SiC, compared to the segregated 
clusters. Though the ``bucky'' clusters  are void in their interior which is not the case for any SiC crystal lattice. We thus expect a transition 
from cage- to bulk-like at some not further specified size n $\ge$ 16.

\subsection{Gas conditions in the circumstellar envelope}

\begin{table}[h!]
\caption{Overview of the SiC nucleation by monomer addition at different temperatures T (in K) and pressures p (in dyne$\cdot$cm$^{-2}$). A 
energetically 
feasible nucleation is marked with $\surd$. Suppressed nucleation with large energy barriers is marked with $\times$. If the nucleation pathway is 
partially favourable, the largest preferential cluster is given.\label{feasabil}}
\begin{tabular}{ l | c | c | c | c | c | c}
T / p &   500 & 100 & 10 & 1 & 0.1 & 0.01\\
\hline
5000 & $\times$ & $\times$ & $\times$ & $\times$ & $\times$ & $\times$\\
3000 & $\times$ & $\times$ & $\times$ & $\times$ & $\times$ & $\times$\\
2500 & Si$_{3}$C$_3$ & Si$_{3}$C$_3$ & Si$_{2}$C$_2$ & $\times$ & $\times$ & $\times$\\
2000 & $\surd$ & $\surd$ & Si$_{3}$C$_{3}$ & Si$_{2}$C$_{2}$& Si$_{2}$C$_{2}$ & Si$_{2}$C$_{2}$\\
1500 & $\surd$ &  $\surd$ & $\surd$ & $\surd$ & $\surd$ & $\surd$ \\
1000 & $\surd$ & $\surd$ & $\surd$ & $\surd$ & $\surd$ & $\surd$\\
500  &  $\surd$&  $\surd$ &  $\surd$& $\surd$ & $\surd$ & $\surd$ \\

\end{tabular}
\end{table}

In Table \ref{feasabil}, the energetic feasibility  for SiC cluster growth at characteristic circumstellar conditions are displayed. 

The left upper part of Table \ref{feasabil} represents gas conditions shortly after the passage of a pulsational shock, where the 
gas is hot and 
compressed (T=3000-5000 K, p= 100 - 500 dyne$\cdot$cm$^{-2}$ = 10 - 50 P). In this case, the SiC dimerisation, representing the 
initial process to start the 
particle growth, is suppressed by an energy barrier of several eV. 

Also for the larger clusters (n $\ge$ 3) the Gibbs free energy of formation, $\Delta$G, of the 
lowest-lying clusters become largely positive and nucleation is unlikely to occur, owing to the lack of stability and high 
activation barriers. 

\begin{figure}[h!]
\plotone{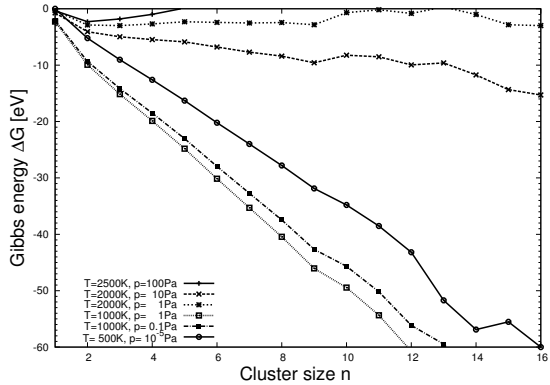}
\caption{Gibbs energy of formation of the ground state clusters versus cluster size for different sets of gas 
temperatures and pressures.\label{Tpenergy}}
\end{figure}

The left intermediate part of Table \ref{feasabil} (T=2000-2500 K, p= 10 - 100 dyne$\cdot$cm$^{-2}$ = 1-10 P) reflects typical 
conditions at the visual 
photosphere where the optical depth $\tau$ is 2/3. Under such conditions, the initial steps for SiC cluster growth are likely to 
occur, as they 
proceed under the excess of energy. At some point of the nucleation chain, however, owing to energy barriers, the growth may not 
proceed (we refer to this as a waiting point)  until the conditions in the wind have relaxed to lower temperatures and densities, where subsequent nucleation is favourable.
Examples for waiting points are Si$_2$C$_2$ and Si$_3$C$_3$, but also also Si$_9$C$_9$ at high temperatures as can be seen in Figure \ref{Tpenergy}. \   

Cooler and more diluted gas conditions prevail further away from the star ($\sim$ 10 R$_*$), where the pulsational shocks have 
strongly weakened and 
damped and the wind has accelerated up to its terminal velocity (and it is assumed that a considerable 
amount of dust has already formed). Such conditions (T=500 K, 10$^{-5}$ dyne$\cdot$cm$^{-2}$ = 10$^{-6}$ Pa) are found in the 
right lower part of 
the table. In this regime, the complete nucleation pathway is energetically favourable. However, the densities are so low, that 
particle collision 
events with subsequent nucleation
 become rare. Nevertheless, previously synthesized dust clusters may stochastically coalesce and 
form dust grains.\    

In summary, SiC cluster formation and growth favours dense and cool conditions; vice versa, the SiC cluster synthesis is hampered in hot and dilute 
environments.
As circumstellar envelopes cover a broad range of temperatures and pressures in space (due to the radial distance from the star) as well as in time 
(owing to dynamical pulsations and wind acceleration), a combination and exposures of various pressures and temperatures involving waiting points is 
more realistic than assuming thermodynamic equilibrium. 
\cite{2012ApJ...745..159Y} showed in their calculations that SiC grains hardly form in local thermodynamic equilibrium (LTE), and that non-equilibrium processes (like pulsations) 
are necessary to explain the observed ratio of SiC dust (0.01-0.3) in carbonaceous dust grains inferred from the radiative transfer model.

For a constant pressure of p=100 dyne$\cdot$cm$^{-2}$ (which corresponds to gas density of 3.6 $\cdot$ 10$^{14}$ cm$^{-3}$) and T=2500 K, the cluster growth is 
energetically feasible up to n=3, or Si$_3$C$_3$. The synthesis of larger cluster sizes is strongly hampered by 
energy barriers of the order of 100 kJ/mol ($\sim$ 1 eV).
At T$\le$2000 K and reasonably high densities, the processes increasing the cluster size are energetically downhill up to the maximum size n=16 
considered in this study.

\begin{figure}[h!]
\plotone{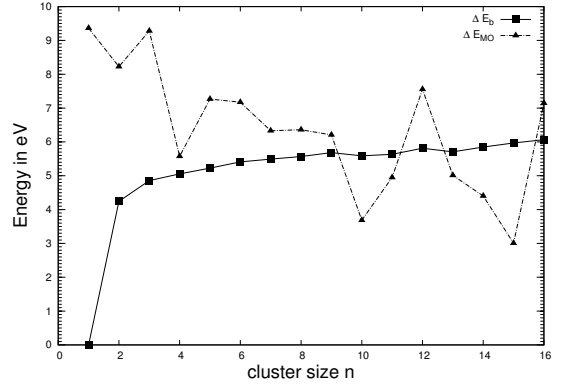}
\caption{The relative binding energy $\Delta E_{b}$ (filled squares and solid line) of the ground state Si$_n$C$_n$ clusters (normalised to cluster size n) and the HOMO-LUMO energy gap $\Delta E_{MO}$ (triangles and dashed line) of the ground state Si$_n$C$_n$ clusters \label{figEdiff}}
\end{figure}

In Figure \ref{figEdiff}, the relative binding energy of the lowest lying Si$_n$C$_n$ cluster (ground state) is plotted versus cluster size $n$ according to the prescription:
\begin{equation}
\Delta E_{b} (Si_{n}C_{n}) = \frac{E_{b}(Si_{n}C_{n})}{n} - E_{b}(SiC)
\end{equation}

The largest incremental in the binding energy (4.3 eV) between clusters of size n and (n+1) occurs between the SiC monomer and the dimer. For larger 
cluster sizes the binding energy increases almost monotonically and saturates around 6.0 eV. 
However, we also note that the ground states of n=9 and n=12 are particularly stable.  
$\Delta E_{MO}$ denotes the energy gap of the highest occupied molecular orbital (HOMO) and the lowest unoccupied 
molecular orbital (LUMO).
This quantity describes the strength and stability of a given electronic configuration. A large $\Delta 
E_{MO}$ indicates a high cluster stability against thermal and radiative excitations. 
Evidently, for $\Delta E_{MO}$, there is no correlation with cluster size n. However, it reveals that some 
cluster sizes (n=3,12,16) have a higher stability and that the closed cage structures are particularly stabel, compared to other cluster sizes. Generally, $\Delta E_{MO}$ tends to decrease the larger 
the system is, as the density of (unoccupied) states increases with cluster size n. 
For a given cluster size, the HOMO-LUMO energy gap is not necessarily the largest for the ground state. 

\begin{figure}[h!]
\plotone{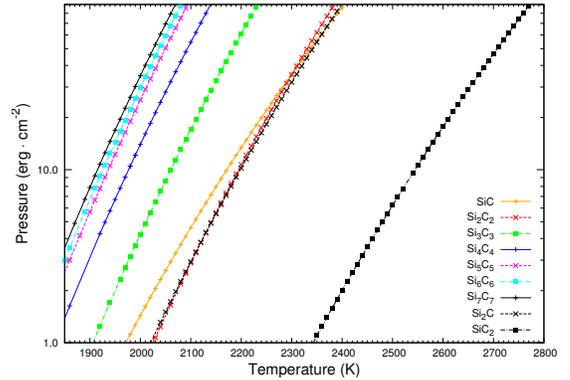}
\caption{Curves with vanishing Gibbs energy of formation, ($\Delta$G$_{f}$=0) for small ground state clusters. 
The corresponding cluster formation is energetically favourable ($\Delta$G$_{f}<$0) for temperatures/pressures 
left/above of these curves. In contrast, for temperatures/pressures right/below these curves, the corresponding 
cluster formation is unlikely ($\Delta$G$_{f}>$0)  \label{Tpgibbs}}
\end{figure}

In Figure \ref{Tpgibbs} curves with vanishing Gibbs energy of formation, ($\Delta$G$_{f}$=0), for small 
Si$_n$C$_n$, n$\le$ 7, and Si$_2$C and SiC$_2$ are shown. Our results indicate that the latter (SiC$_2$) is the first 
silicon-carbon molecule to emerge from the hot atmosphere, as it is more stable over a broad range of gas pressures, 
compared with the other considered compounds. 
The formation of Si$_2$C and Si$_2$C$_2$ becomes exogonic ($\Delta$G$_{f}<$0) at very similar pressures and temperatures.  
In contrast, the formation of the SiC molecule is more likely at lower temperatures assuming a constant pressure.
For the larger Si$_n$C$_n$ clusters, the formation probability shifts successively to lower temperatures (or higher pressures).   
From Figure \ref{Tpgibbs} we conclude that a homogeneous nucleation is viable, 
presuming a bottom-up approach in the formation of Si$_n$C$_n$ clusters. 
It should be noted that, in the presence of stellar pulsations, a trajectory of a gas parcel will not be represented by 
a single line in the T-p diagram, but rather by a complex zigzag. 
It implies that certain clusters with a particular high stability (i.e. waiting points) form at several times during the nucleation process and thus can be considered as candidates for a possible observational detection. 
In summary, we conclude that a homogeneous nucleation of SiC clusters 
is definitely viable in circumstellar environments and that, owing to their thermal stability, the molecular species Si$_2$C$_2$ 
and Si$_2$C may play an important role in the initial steps of SiC nucleation.

\begin{figure}[h!]
\plotone{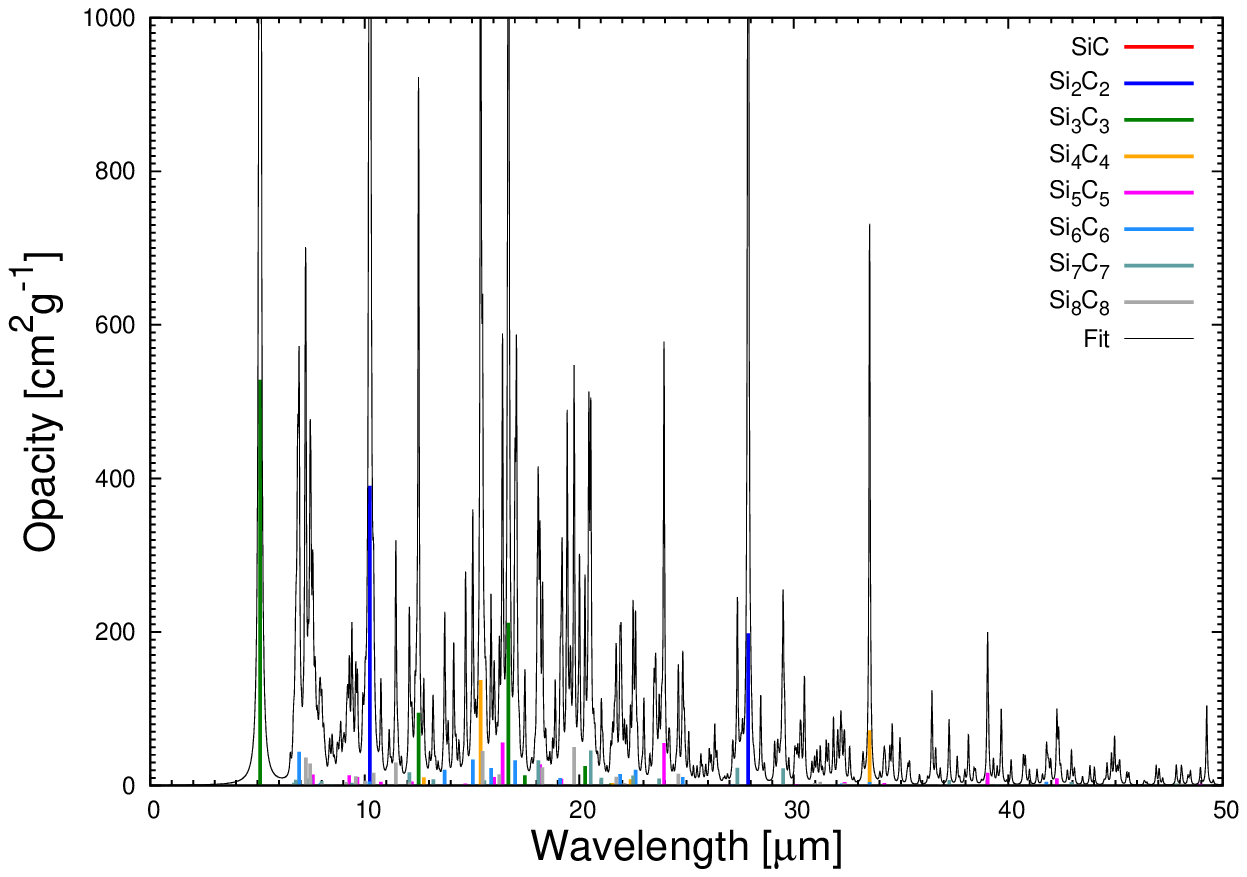}
\plotone{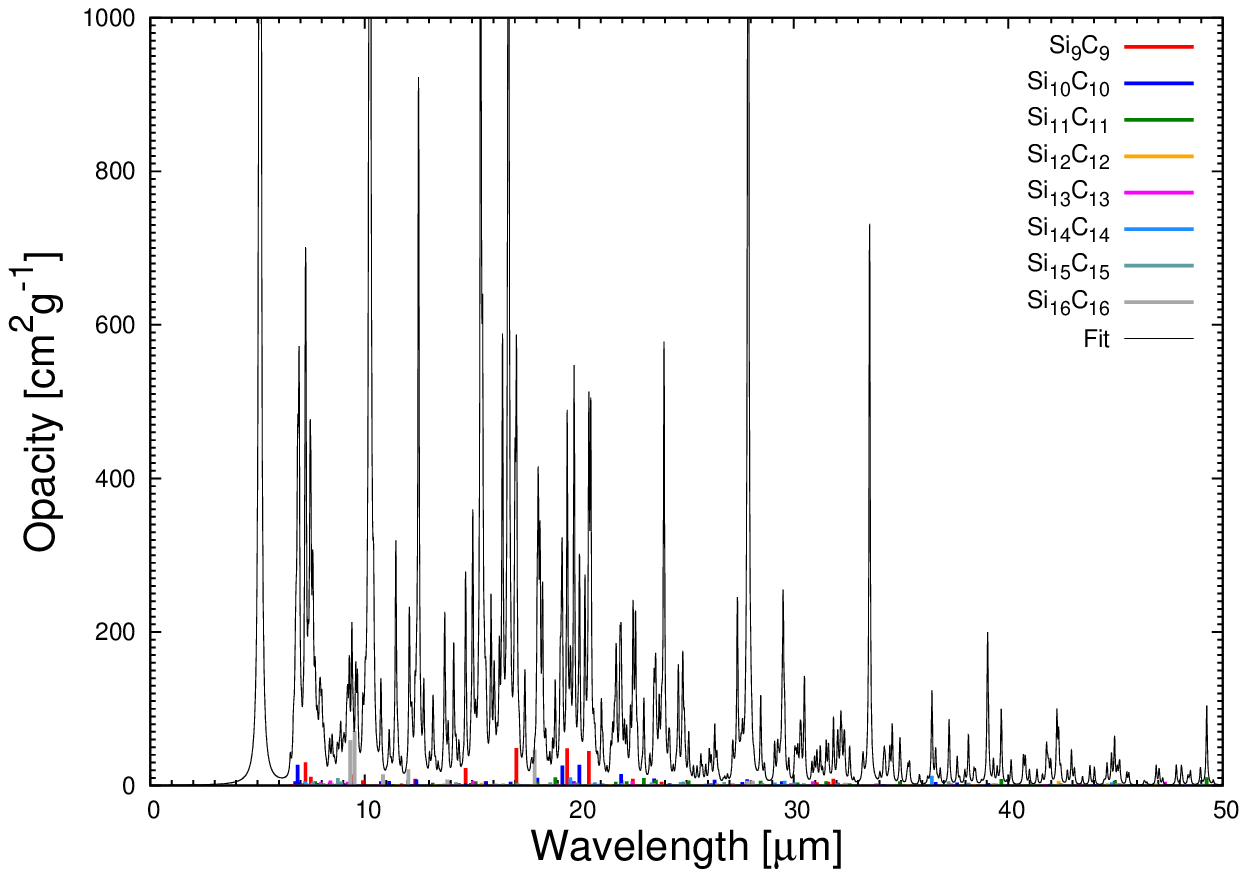}
\caption{Vibrational spectra of the ground state Si$_n$C$_n$ clusters. The fit is a Lorentzian distribution with 
a FWHM parameter of 0.033.  
\label{figIRspectra}}
\end{figure}

In Figure \ref{figIRspectra}, the calculated vibrational IR spectra of the Si$_{n}$C$_{n}$ ground state clusters 
are displayed.
A cluster of size n (i.e. the number of SiC units) exhibits at maximum 6$\cdot$(n-1) individual vibrations 
accounting for bending and streching 
modes. 
Clusters with symmetric arrangement, however, show fewer emission modes, owing to the multiplicity of identical 
vibrations.

The unit conversion from km mole$^{-1}$ to the opacity in cm$^2$g$^{-1}$ is obtained by

\begin{equation}
 1 \rm{km\hspace{0.1cm}mole^{-1} = \frac{10^{5}}{\overline{\nu}} cm^2 mole^{-1} = \frac{10^{5}}{\overline{\nu }M} cm^2 g^{-1}}
\end{equation}

\noindent where $\overline{\nu}$ is the vibrational frequency in units of cm$^{-1}$ and M the molecular mass in atomic mass units (e.g. $\sim$ 40 for the SiC 
monomer).

\section{Discussion}

The 11.2-11.5 $\mu$m feature observed in the spectra of carbon-rich stars represents the most prominent emission attributed to SiC dust particles.
About 4 \% of the stars in the sample of \cite{1986ApJ...307L..15L} show a feature shifted to 11.6 $\mu$m corresponding to the wavelength of the SiC 
molecule (monomer) vibration mode. Some carbon stars exhibit a second peak around 11.7 $\mu$m \citep{1995ApJ...449..246G}. These stars are characterised by a lower feature strength and broadened emission indicating larger SiC particles.
A spectral feature at 9 $\mu$m that correlates with trends of the 11.3 $\mu$m is observed as well in carbon stars \citep{2005ApJ...634..426S}.
The authors concluded that the carrier of the 9 $\mu$m is either amorphous SiC or Si-doped nanodiamond. 
Moreover, the authors find that as the star evolves and increases its mass loss rate, the SiC dust grain sizes become smaller. Finally, in the 
superwind phase, owing to the high mass loss rate, the SiC feature appears in absorption, broadens, weakens, and is shifted towards shorter 
wavelengths.  
\cite{2009ApJ...691.1202S} investigated spectral features in the 10-13 $\mu$m range in a sample of extreme carbon stars and attributed them to 
carbonaceous solids including a fraction of SiC dust. In our study, Si$_2$C$_2$ at 10.237 $\mu$m and Si$_3$C$_3$ at 12.507 $\mu$m show the strongest 
emission in this range among the Si$_n$C$_n$ cluster ground states.
\cite{2015A&A...583A.106R} showed that, apart from the spectral feature around 11.4 $\mu$m, amorphous carbon and SiC dust particles exhibit 
absorption distributions that are fairly similar in the small particle limit. This feature is thus a unique and distinct tracer for the presence of SiC 
dust grains.\
Also laboratory spectra show a wide variety of the SiC phonon features in the 10-13 $\mu$m wavelength range, both in peak wavelength and band shape 
\citep{1999A&A...345..187M}. It is the only relatively broad band that is attributed to SiC. As previously mentioned the SiC crystal type ($\alpha$ 
vs. $\beta$ SiC) plays a minor role in the 10-13 $\mu$m emissivity.\

In our study, we found several clusters with vibrational emissions in the 11.2-11.5 $\mu$m wavelength. Their overall IR intensity is, however, too 
small in order to explain the observed emission. The investigated cluster sizes (up to n=16) may be too small to reproduce the bulk-related phonon 
emission around 11.3 -11.4 $\mu$m. 
The SiC molecule (monomer) exhibits a vibrational emission feature at 11.599 $\mu$m. However, its IR Intensity is weak (0.1274 km$\cdot$mole$^{-1}$ = 
0.3694 cm$^2$g$^{-1}$), compared with the IR intensities of the other SiC clusters.  
Also the three-atomic species Si$_2$C and SiC$_2$, the latter being a by-product of SiC dust formation, cannot account for the 11.3 $\mu$m feature 
in their spectra. 
Si$_3$C$_3$ (isomer F) exhibits a feature at 11.325 $\mu$m with a reasonable intensity of 106.2789 km$\cdot$mole$^{-1}$ = 100.3 cm$^2$g$^{-1}$. However, 
the cluster is 0.84 eV above the ground state at standard conditions, and the situation is similar (0.8-1.2 eV) at characteristic wind conditions.
Si$_4$C$_4$ (isomer F) shows at 11.298 $\mu$m signature with a strength of 85.7498 km$\cdot$mole$^{-1}$ = 60.5501 cm$^2$g$^{-1}$ and has a potential energy 
of 0.4-0.6 eV above the ground state, depending on the gas conditions.
The ground state of Si$_5$C$_5$ (isomer A) shows an emission at 11.298 $\mu$m, but the IR intensity (0.2336 km$\cdot$mole$^{-1}$) is (too) low.
Other Si$_5$C$_5$ isomers show signatures in this wavelength range, namely, structure C (which is the minimum structure in the higher pressure 
cases) at 11.405 $\mu$m with 3.1128 km$\cdot$mole$^{-1}$, D at 11.424 $\mu$m and 
3.3497 km$\cdot$mole$^{-1}$, E at 11.350 $\mu$m with 0.5734 km$\cdot$mole$^{-1}$, and F at 11.299 $\mu$m with 3.5731 km$\cdot$mole$^{-1}$. Although Si$_5$C$_5$ has several low-lying 
candidate carriers of 11.3 $\mu$m features, the IR intensities are very low, compared with the other spectral features these isomers have.  
For the larger clusters, we compiled a table with vibrational intensities in the 11.2-11.5 $\mu$m range (see Table \ref{11p3}).  

\begin{table}[h!]
\caption{Vibrational emission of the presented clusters (n $\ge$ 6) in the 11.2-11.5 $\mu$m wavelength range. The two columns identifies cluster 
size n and state X, the third column displays the wavelength $\lambda$ in $\mu$m and the intensity I (in km$\cdot$mole$^{-1}$ and cm$^2$g$^{-1}$) is listed 
in the fourth column.\label{11p3}}
\begin{tabular}{ c c|| r | r | r}
n & X & $\lambda$ (in $\mu$m) & I (in km$\cdot$mole$^{-1}$) & I (in cm$^2$g$^{-1}$) \\
\hline
6 & B & 11.365 & 1.0139 & 0.4801 \\
7 & B & 11.510  & 14.8496 & 6.1042\\
8 & A & 11.449  & 83.6211 & 29.9181\\ 
10 & A & 11.455  & 9.9048 & 2.8365\\
11 & A  & 11.384  & 8.1423 & 2.1066\\
11 & D  & 11.181  & 5.1542 & 1.3098\\
12 & C & 11.424  & 3.9820 & 0.9477\\
12 & D & 11.461  & 7.4771 & 1.7853 \\
13 & B &  11.449  & 3.4249 & 0.7541\\
13 & C  & 11.293 & 0.3679 & 0.080\\
13 & D & 11.219 & 24.8402 & 4.9765\\
14 & C & 11.363 & 0.8570 & 0.1739\\
   &   & 11.289 & 2.8705 & 0.5787\\   
15 & B & 11.227 & 7.8250 & 1.4641\\
15 & C & 11.213 & 4.8333 & 0.9032\\
16 & C & 11.363 & 5.7596 & 1.0226\\
\end{tabular}
\end{table}

For cluster size n=8,11,12,13 and 15 the ground states (or next higher lying states) emit in this wavelength regime; the intensities are (apart from 8A), however, quite low. Large abundances of a specific cluster, though, could increase the intensity significantly. \\

Some of the spectral peaks identified in our calculations are not explicitly reported in the literature. Nevertheless,  SiC clusters may represent a 
key player for the onset of dust formation in carbon-rich AGB stars, albeit not directly detected. 

In fact, \cite{1989Natur.339..196F} found experimental evidence for a scenario in which SiC nucleates at higher temperatures and provides surfaces 
for subsequent carbon condensation in a hydrogen-rich atmosphere. 
Moreover, \cite{1994ApJ...429..285C} used the model of induced nucleation, where the grain growth proceeds on reactive surfaces of pre-existing seed particles, and showed that a subsequent condensation of carbonaceous material results in composite grains that are consistent with grains found in pristine meteorites.
\cite{1996A&A...307..551K} thus suggested that SiC grains form at high temperatures by homogeneous nucleation, but as soon as the temperature has decreased (i.e. at larger radii), a mantle of amorphous carbon (amc) may deposit on SiC seeds. 
The spectral signatures of pure SiC may thus be blended and/or suppressed by the amc mantle.

There is a series of UnIdentified Bands (UIBs) at 3.3, 6.2, 7.7, 8.6 and 12.7 $\mu$m, respectively, seen in carbon-rich AGB stars. These features are commonly attributed to Polycyclic Aromatic Hydrocarbon (PAH) emission \citep{1996A&A...315L.369B, 1998A&A...335L..69H, 2000A&A...356..253J, 2006A&A...447..213B}. 
As both PAHs and SiC clusters,  contain aromatic C$_6$ rings and have conjugated bonds in common, they may show a remarkable spectral similarity.
In the following we examine whether SiC clusters could account for the emission of UIBs or not.  For the 3.3 and 6.2 $\mu$m band we find no coincidence with the vibrational spectra of SiC clusters. Around 7.7 $\mu$m the SiC ground state clusters with n =11,13,14 and 16 show emission.
We find that the ground states Si$_{15}$C$_{15}$ and Si$_{16}$C$_{16}$ emit at 8.6 $\mu$m.  Around 12.7, we find vibrational modes of the n=14 and n=15 ground states. All these structures have C$_6$ and C$_5$ rings in common. The symmetric ``bucky''-like structures B in Figure \ref{fig12} and structure A in Figure \ref{fig14} do not exhibit spectral features at these wavelengths.

Next, we want to address the viability of SiC cluster nucleation and the derivation of (parametrised) reaction rates from the analysis of our results. 
So far, we identify the most likely cluster structures and pathways in SiC nucleation and dust formation.
Rate determination is though difficult to achieve, as it requires knowledge about the (various) transition states involved in the reaction. 
A directly proceeding reaction could be evaluated, as it depends only on the (calculated) properties of reactants and the products.
Such reactions are, however, unlikely to occur,
in particular, for the gas-phase chemical reactions of SiC monomers and dimers representing the crucial starting point in the present bottom-up approach.
Unfortunately, these rates are poorly characterised. Only two chemical reactions rates for the SiC monomer are reported in NIST
\citep{Nist2005}:

\begin{itemize}
\item  Si + C $\rightarrow$ SiC + h$\nu$ (radiative association)
\item  Si + CH$_2$ $\rightarrow$ SiC + CH (bimoecular collision)
\end{itemize}

\noindent whereas the bimolecular collision reaction is estimated by analogy to the reaction Si + CH$_3$ $\rightarrow$ SiCH + H$_2$ \citep{KIN:KIN1071}. 
Despite is low energy barrier of 136.73 K (1.14kJ/mol) the radiative association reaction is very slow \citep{2009MNRAS.400.1892A}.
Moreover, owing to the lack of gas phase reaction rates, isovalences of Si and C are presumed, and rates for SiC are equalized with rates for 
C$_{2}$ (see e.g. \cite{2010ApJ...713....1C}). This may be adequate as a first approximation.
However, the binding energy of SiC (4.71 eV) is higher by more than 1 eV compared with C$_2$ (3.6 eV). 
Moreover, the Si-C bonding has a small, but not negligible dipole moment of $\sim$ (1.7-1.8) Debye, due to the larger 
size and the higher number of electrons of the Si atom. This may have non negligible effects on the reactivity of the molecules.\

\cite{2012ApJ...745..159Y} provide reaction enthalpies for SiC cluster growth for temperatures 1500 K and 1000 K. 
The enthalpies indicate that a homo-molecular cluster growth (i.e. the addition of SiC molecules to a Si$_n$C$_n$ cluster) is the energetically most favourable formation route. 
Albeit the reaction enthalpies are approximated with that of solid SiC for n \textgreater 3, they conclude that the reactions

\begin{equation}
Si_{n}C_{n} +SiC \rightarrow Si_{n+1}C_{n+1}
\label{homog}
\end{equation}

are the dominant processes in the formation of SiC dust grains, consistent with our findings (see Figure \ref{Tpenergy}).\\

In the following, we list observation of silicon carbon molecules in C-rich AGB stars and compare them with our findings.
The molecular SiC radical has been detected first in CW Leo by \cite{1989ApJ...341L..25C}.
We find a rotational constant of 20643.1 MHz consistent with the spectroscopic constant B=20297.6 MHz.
Note, that SiC is a triplet and thus the rotational level is split into three states.
Our M11 calculations of Si$_2$C yield the following rotational constants A=58363.8 MHz, B=4567.1 MHz and C=4235.7 MHz, whereas the derived constants in \cite{2015ApJ...806L...3C} (S reduction) as A=64074.3, B=4395.5, C=4102.1 are slightly different, but still compatible.
For SiC$_2$, we obtain the following rotational constants A=53511.7 MHz, B=13004.6 MHz and C=10462.1 which are in good agreement with the laboratory (A=53909 MHz, B=13530 MHz, C=10751 MHz) and observational data (A=52390 MHz, B=13156.2 MHz, C=10447.4 MHz) of \cite{1984ApJ...283L..45T}.
The most stable isomer of SiC$_3$ has a cyclic geometry and was detected in CW Leo (\cite{1538-4357-516-2-L103,2000A&AS..142..181C}). We find 
rotational constants of 39.962 GHz and 6.240 GHz with the M11 functional which is relatively close to the laboratory spectra of 37.9 and 5.83 GHz, respectively.
Linear SiC$_4$ was detected by \cite{1989ApJ...345L..83O} in CW Leo.
With the M11 functional we obtain a rotational constant of 1549.6 MHz, close to the observed value of 1533.8 MHz.\

Recent observations revealed that, among the silicon carbon molecules, SiC$_2$ and Si$_2$C dominate the inner envelope whereas the SiC molecule is 
2-3 orders of magnitude less abundant \citep{2015ApJ...806L...3C}.
We conclude that the SiC molecule is rapidly converted in SiC$_2$, Si$_2$C and Si$_n$C$_n$ clusters.
The emission of SiC$_3$ and SiC$_4$ arises in the intermediate and outer envelope of CW Leo. Thus, we suggest that the latter molecules are the 
result of photochemistry or grain surface reactions and that they do not play a role in the nucleation of SiC dust.
Assuming a dust-to-gas mass ratio of 2.5 $\cdot$ 10$^{-3}$ and a fraction of 10 \% SiC in the dust grains results in a solid SiC abundance, 
(SiC)$_{dust}$/H$_2$, of 1.25 $\cdot$ 10$^{-5}$. This is slightly less the half of the solar Si abundance (3 $\cdot$ 10$^{-5}$).
It has been suggested that the recently discovered Si$_2$C molecule plays a key role in the formation of SiC dust grains 
\citep{2015ApJ...806L...3C}. Although the molecules is well characterised in terms of 
geometry and energetics, reaction rates are lacking for Si$_2$C. 
In our study, we find that the Gibbs free energy of formation of SiC$_2$ is lower by at least 100 kJ/mol compared 
to Si$_2$C for all p-T combinations listed in Table \ref{feasabil}. In fact, the latter explains the observed higher 
SiC$_{2}$ abundance between 1 and 4 R$_{*}$, compared with Si$_2$C. In the intermediate envelope region (4-40  R$_{*}$) equal amounts of Si$_2$C and SiC$_2$ are present. Further out ($\sim$ 40-1000 R$_{*}$), again  SiC$_{2}$ is favoured over Si$_2$C, before both species are essentiall dissociated/depleted.
These results indicate that SiC$_2$ is favoured over (or at least equivalent to) Si$_2$C in circumstellar outflows, and agrees with our 
calculations, assuming a formation pathway via the SiC molecule and equal amounts of Si and C atoms. 
As carbon is $\sim$ 17 times more abundant than silicon, assuming scaled-solar abundances, the dominance of SiC$_2$ versus SiC and Si$_2$C is even 
emphasized. Owing to the excess of carbon relative to silicon, the molecular species Si$_3$C and Si$_4$C are excluded from the present study.
In carbon-rich atmospheres of evolved AGB stars, this (C/Si) ratio tends to be even higher and has values $\sim$ 20-30 \citep{2015ApJS..219...40C}.\
As can be evaluated from Table \ref{delHrxn} cluster growth via  SiC$_2$ is energetically unfavourable at temperatures of 1000 K and 1500 K.
The formation of SiC$_2$ represents thus a competing branching to the synthesis of Si$_n$C$_n$, n $\ge$ 3 clusters for conditions close to the 
star. Therefore, we conclude that SiC$_2$ is a by-product of SiC dust formation in the inner envelope. Further away from the star, at lower 
temperatures and pressures, however, the cluster nucleation via SiC$_2$ becomes exothermic and exergonic and thus also likely to occur. 
A nucleation pathway involving Si$_2$C as intermediary is energetically thoroughly viable, also close to the star, albeit a 
cluster growth according to Equation \ref{homog} is expected to be faster and more efficient. Our theoretical findings thus explain and
reflect the observed radial abundance profiles of the silicon carbon molecules in CW Leo.\\

In Table \ref{delHrxn} we compare the energetics of the ground-state clusters derived in this study with \cite{2012ApJ...745..159Y}. 
The authors evaluated the reaction enthalpies $\Delta$H$^0$ from a data set by \cite{Deng2008} at 1000 and 1500 K. We find similar trends in the 
exothermicity of the reactions, though our value are systematically lower by 3-77 (16-86) kJ/mol for T=1000 K and 24-108 kJ/mol 
for T=1500 K.	

\begin{table}[h!]
\caption{Reaction enthalpy $\Delta$H$^0$ in kJ$\cdot$mole$^{-1}$ for SiC cluster growth reactions and comparison with the studies of \cite{2012ApJ...745..159Y,Deng2008} at 
temperatures of 1000 K and 1500 K.\label{delHrxn}}
\begin{tabular}{ l | c  c | c  c}
  & \multicolumn{2}{c|}{our study} & \multicolumn{2}{c}{Yasuda+(2012) study} \\ 
  & \multicolumn{2}{c|}{$\Delta$H$^0$} & \multicolumn{2}{c}{$\Delta$H$^0$} \\
N  & 1000 K & 1500K & 1000K & 1500K \\
\hline
R1 & -807.8 (-821.1) & -804.5 (-816.9) & -751.5 & -751.0 \\
R2 & -572.6 (-605.0) & -566.9 (-623.0) & -518.8 & -514.5 \\
R3 & 15.1 (9.7) & 19.57 (15.1) & 77.0 & 78.1 \\
R4 & 250.3 (225.7) & 257.2 (209.0) & 309.7 & 314.7 \\
R9 & -187.1 (-134.6) & -183.2 (-133.9) & -110.8 & -109.8 \\
R10 & 48.2 (81.5) & 54.4 (55.0) & 121.9 & 126.7 \\
R11 & 235.3 (216.1) & 237.6 (183.9) & 232.7 &  236.5 \\

\end{tabular}
\vspace*{1cm}
\newline
where
\vspace*{1cm}
\begin{tabular}{ l | l}
 R1 & SiC + SiC $\rightarrow$ Si$_2$C$_2$\\
 R2 & SiC + Si$_2$C$_2$ $\rightarrow$ Si$_3$C$_3$\\
 R3 & SiC + SiC$_2$ $\rightarrow$ Si$_2$C$_2$ + C\\
 R4 & SiC$_2$ + Si$_2$C$_2$ $\rightarrow$ Si$_3$C$_3$ + C\\
 R9 & Si$_2$C + SiC $\rightarrow$ Si$_2$C$_2$ + Si\\
 R10 & Si$_2$C + Si$_2$C$_2$ $\rightarrow$ Si$_3$C$_3$ + Si\\ 
 R11 & Si$_2$C$_2$ + Si$_2$C$_2$ $\rightarrow$ Si$_3$C$_3$ + SiC\\
\end{tabular}
\end{table}

Further listed reactions in \cite{2012ApJ...745..159Y} could not be compared, as we have not investigated the species SiC$_3$, SiC$_4$, 
Si$_2$C$_3$, Si$_3$C, Si$_3$C$_2$, Si$_4$C, Si$_4$C$_2$ and Si$_5$C.
The reason for the systematic offset may arise due to the use of different functionals / basis sets in \cite{Deng2008} (M11/cc-pVTZ and B3LYP/cc-pVTZ vs. 
B3PW91/6-31G(d)), other ground state clusters and by the use of combined thermo-chemistry databases.\\

Another point that we aim to address is the cluster physics. Therefore, we compare our results to studies of SiO - representing a counterpart to SiC in oxygen-rich environments - and TiC, another metallic carbide.

Silicon oxide (SiO) is a key ingredient for the formation of the astronomcally relevant and abundant silicates of pyroxene and olivine.
Under circumstellar conditions, homogeneous SiO nucleation is limited by considerably large energy barriers of the order $\sim$ 1 eV \citep{2012MNRAS.420.3344G,C6CP03629E}. 
Moreover, the (SiO)$_n$ global minimum structures show segregations in the form of Si-Si bonds for sizes n$>$5.
In contrast to silicon oxide clusters, (SiO)$_n$, the most energetically favourable SiC clusters tend to have alternating Si-C bonds for sizes larger than 12 units and exhibit segregations for sizes n$<$12. We thus observe opposing trends in the degree of segregation versus size for SiO and SiC clusters. 
Furthermore, homogeneous SiC nucleation is feasible in cirumstellar environments and may occur even at elevated temperatures (T=2000K). 
Titium carbide (TiC) is found in the centers of pristine meteoric grains and laboratory measurements of small-sized TiC nano-crystals show a prominent spectral feature around 21 $\mu$m \citep{2000Sci...288..313V}. 
However, \cite{2003ApJ...587..771C} demonstrated that TiC grains are implausible carriers of the observed infrared 21 $\mu$m feature around carbon-rich post-AGB stars. 
Recent investigations of small TiC$)_n$ (n=6,12) clusters have shown that the lowest energy structures possess a cubic geometry with alternating Ti-C bondings \citep{Lamiel-Garcia2014}. 
Isomers deviating form pure alternating bonds (i. e. exhibting C-C bonds) have potential energies slightly above the cubic forms. 
We thus conclude that segregation plays a negligible (or minor) role in homogeneous TiC nucleation and that the transition to the crystalline bulk material takes place at comparable small sizes.

\section{Summary}
We have found energetically favourable clusters for (SiC)$_n$ up to a size of n=16. 
The results are used to predict the viability of nucleation and the reaction probability in SiC cluster chemistry. 
Our findings show that SiC dust formation is viable in the dense cooling atmospheric gas layers by addition of 
single SiC gas phase molecules 
(homogeneous 
nucleation). The nucleation pathway includes waiting points, where the SiC addition may be energetically 
unfavourable. Nevertheless, nucleation  
owing to changes in gas conditions (e.g. shocks, radiation) is not unlikely.  
The 11.3 $\mu$m feature represents an emission which is uniquely attributable to SiC dust grains in the near 
infrared regime. There is a number of clusters showing emission around this feature. However, their overall intensities are rather 
low. We thus conclude that the major contribution to 11.3 $\mu$m emission arises from bulk SiC material.
The rate enthalpies compare qualitatively well with a previous study \citep{2012ApJ...745..159Y}, albeit the data 
derived by our study is
systematically lower by $\sim$ 3-77 (16-86) kJ/mol.

We acknowledge the CINECA award under the ISCRA initiative, for the availability of high performance computing 
resources and support. 
The authors gratefully acknowledge the referee whose constructive comments greatly helped improving the quality of the present paper. 
We acknowledge L. Decin for useful discussions on the comparisons with observations. 
The figures displaying cluster structures have been produced with the help of the program MOLDEN 
\citep{Schaftenaar2000}. All other Figures have been created with gnuplot.

\end{document}